\documentstyle[prd,aps,preprint,tighten,epsf]{revtex}

\newcommand{\pslash}[1]{\rlap{/}\kern-0.8pt #1}
\newcommand{\lslash}{\rlap{/}\kern-0.0pt l}
\newcommand{\Dslash}{\rlap{/}\kern-2.0pt D}

\begin{document}
\bibliographystyle{apsrev}
\preprint{CU-TP-1000}
\title{Domain Wall Fermion Study of Scaling in Non-perturbative 
Renormalization of Quark Bilinears and $B_K$}
\author{Y.~Zhestkov}
\address{Physics Department,
Columbia University,
New York, NY 10027}
\date{February 8, 2001}
\maketitle

\begin{abstract}
We compute non--perturbatively the renormalization coefficients
of scalar and pseudoscalar operators, local vector and axial currents,
conserved vector and axial currents, and $O_{LL}^{\Delta S=2}$
over a wide range of energy
scales using a scaling technique that connects the results of
simulations at different values of coupling $\beta$. We use the domain
wall fermion formulation in the quenched approximation at a series of
three values of $\beta$, 6.0, 6.45, and 7.05, corresponding to lattice
spacing scaling by factors of two.  
\end{abstract}

\section{Introduction}
\label{sec:intro}

Lattice QCD has proven to be a powerful approach to calculating from the
first principles the mass spectrum, weak decay constants, weak matrix
elements, quark masses, and hadronic structure
functions. Such studies require calculating 
matrix elements of composite operators. In order to extract physical
predictions in the continuum limit from Monte--Carlo simulations
in lattice QCD, one in general needs to 
determine the normalization of these operators in terms
of physical requirements.

In calculations involving the operator product expansion one needs to
know the Wilson coefficients. The corresponding calculations are done
perturbatively, and therefore must be carried out at a
 high enough energy scale.
To evolve those coefficients down to the scale at which lattice calculations
are performed, one usually also uses perturbative methods.

A possible approach to renormalization of operators defined on the 
lattice is lattice perturbation theory. This approach, though, suffers from
many difficulties, the main being poor convergence of the perturbation
series. One source of this difficulty was identified by Parisi and
Toe \cite{Parisi:1980},
and later by Lepage and
Mackenzie \cite{Lepage:1993xa},
 and is due to the presence of tadpole diagrams. The authors of
\cite{Lepage:1993xa} proposed
a method to improve the convergence properties of these series
which is known as tadpole--improved perturbation theory.
Nevertheless, the calculations using this method are complex, and therefore
are rarely carried out beyond the one--loop level. Since $g^2$ decreases
only logarithmically with decreasing $a$, the systematic improvement
of perturbation theory is very difficult, leaving the possibility
for large systematic uncertainty in extraction of physical results.

In some limited cases renormalization of lattice operators can be
obtained from Monte--Carlo simulations using the chiral Ward Identities.
The examples include vector and axial vector currents, and the ratio of
pseudoscalar and scalar densities.
Unfortunately, this method cannot be applied to general 
composite operators for which there is no corresponding Ward Identity.

A method for full non--perturbative renormalization of lattice operators
by means of Monte--Carlo simulations was proposed in 
\cite{Martinelli:1995ty}.
The procedure mimics the approach taken in continuum perturbation theory.
The renormalization conditions are imposed directly on quark and gluon
Green's functions, in a fixed gauge, with specified, external lines
carrying far off--shell momenta. This defines the so--called Regularization 
Independent scheme (RI). At high momenta the results can be 
directly related to the 
perturbative calculations, provided that corresponding conversions are
made in order to match different renormalization schemes used in continuum
calculations (usually the $\overline{MS}$ scheme).

A necessary condition for the RI method to work is to have the
off--shell momentum scale of the external lines, $\mu$,
much smaller than the inverse lattice
spacing and much larger than $\Lambda_{QCD}$. The former condition
restricts the value of reliable momenta from above for a given lattice
spacing $a$. The lower bound on momentum comes from the
need to compare with perturbation theory which is only applicable
at an energy scale above the non--perturbative regime, 
$\mu\gg\Lambda_{QCD}$.
These two constraints leave a pretty narrow
range of momenta that can be used to calculate renormalization coefficients.

To extend the range of momenta over which an operator is renormalized,
one can perform simulations at different values of 
coupling constant $\beta=6/g^2$ with
overlapping regions of reliable physical momenta. By combining the results
of these calculations, it is in principle possible to relate renormalization
coefficients of a given operator between 
quite different momentum scales $\mu$.
By starting with sufficiently high momenta where the perturbation theory can
be trusted, operators can be renormalized at a series of scales 
down to a few GeV where most lattice calculations are done.

To make the matching between results obtained at different $\beta$'s
meaningful, it is important to insure that it is done under 
identical physical
conditions. In particular, mass and volume dependence have to be taken into
account.

The difficulty that arises from the finite volume effects can be approached
in several ways. A direct way of taking the infinite volume limit turns
out to be inefficient due to the strong volume dependence 
 at low momenta. In this work we instead keep the volume explicitly
finite. The matching between calculations at different $\beta$'s is done at the
same physical momentum and physical volume. This procedure is described
in detail in Section \ref{sec:Scaling}.

In our simulations we use the domain wall fermion formulation
\cite{Kaplan:Kaplan}--
\cite{Vranas:2000tz}. This approach is computationally more demanding
than the usual Wilson fermions, but offers significant
advantages that are more than enough to compensate for the additional
computer time required for the calculations. The DWF formulation
provides an easy procedure of taking the chiral limit, independently
of the continuum limit. No additional fine--tuning is required.
 By working at 
sufficiently large lattice extent in the fifth direction, the residual
quark mass arising from the chiral symmetry breaking coupling
between the walls can be made very small. 
Then the limit of zero bare quark 
mass corresponds to the chiral limit with high accuracy.
We use this property to eliminate the mass dependence of our results,
by taking the chiral limit.

 The excellent chiral properties
of the DWF formulation suppresses the mixing of such operators as 
$O^{\Delta S=2}_{LL}$ with operators of opposite chiralities, 
Ref.~\cite{Dawson:2000kh}.
This is a serious complication for studying the renormalization of
such operators using fermion actions with large chiral
symmetry violations. 

It is important that in the DWF formulation
the chiral properties are improved not only for the on--shell
calculations, but also for the off--shell ones.
 This is one of the
reasons why the DWF formulation is ideal for using the non--perturbative
renormalization method described in the next section that imposes
a renormalization condition on the matrix elements of operators
between external off--shell quark states.
Another important consequence of these good chiral properties 
of the lattice theory with the DWF is the absence of order $O(a)$
errors in the results for the matrix elements of operators.

We use the quenched version of the DWF method in our paper. For an
extensive analysis of the properties of this formulation
 the reader is referred to
the early works on this subject, Refs.~\cite{Blum:2000kn} and 
\cite{AliKhan:2000iv}.

The paper is organized as follows: In Section \ref{sec:NPR}
 we explain the method of
non-perturbative renormalization. The scaling technique is described in
Section \ref{sec:Scaling}. In Section \ref{sec:Results}
 we present the results of
numerical calculations.

\section{The Non--Perturbative Renormalization Method\label{secNRI}}

\label{sec:NPR}

In our calculation of non--perturbative renormalization coefficients
we follow closely the procedure first proposed by 
the Rome--Southampton group \cite{Martinelli:1995ty}. 
The renormalized operator $O(\mu)$ is defined as the bare
operator $O_{bare}$ multiplied by the $Z$--factor,

\begin{equation}\label{eq1}
      O(\mu) = Z_O(\mu; a) O_{bare}(a) \;.
\end{equation}

For simplicity of presentation we ignore the possibility of operator 
mixing which would require replacing Eq.~(\ref{eq1}) by a matrix
equation but have no other effect on the method developed in this paper.
Here the bare operator $O_{bare}(a)$ is regularized by discretizing
it on a space--time lattice with lattice spacing $a$ that has
units of inverse energy and throughout this paper $a$ is assumed to
be uniform in all directions. It can also depend on other (bare) parameters
in the Lagrangian. The renormalized operator $O(\mu)$
has low energy matrix elements with no
dependence on regularization parameters.
These low energy matrix elements are functions
of only physical parameters, such as physical masses,
coupling constant, etc., defined at the renormalization 
scale $\mu$, and $\mu$ itself.

The renormalization condition imposed on an operator 
constructed from a product of $n$ quark fields is given by

\begin{equation}\label{rencond}
  Z_O(\mu; a) Z_q^{-n/2}(\mu; a) \Gamma_O(p;a) |_{p^2=\mu^2} = 1 \;,
\end{equation}
where $Z_q(\mu; a)$ is the quark field renormalization coefficient,
\begin{equation}
q_{ren}(\mu) = Z_q^{1/2}(\mu; a)q_{bare}(a) \;.
\end{equation}
This specifies our conventions for $Z_q$. The precise prescription for
determining $Z_q$ is given later in Eq.~(\ref{Zqdef}).

The quantity $\Gamma_O(p)$ is obtained from the amputated matrix element of
the operator $O_{bare}$ between external off-shell
(bare) quark states, in the Landau gauge,
by tracing it with a projector $\hat{P}_O$
on a tree--level operator. In the case of quark bilinear operators it
is defined by the equation

\begin{equation}\label{project}
  \Gamma_O(p;a) = 
        {{\mbox{Tr}\left(\Lambda_O(p;a) \hat{P}_O\right)} 
         \over{12}} \;,
\end{equation}
where
\begin{equation}
  \Lambda_O(p;a) = S_{bare}(p;a)^{-1} G_O(p;a) S_{bare}(p;a)^{-1}
\end{equation}
is the truncated Green's function. 
The corresponding equations for the case of four quark operators 
can be found in Section \ref{sec:bk}  of this paper.

The above renormalization condition is independent of the regularization
scheme, although it does depend on a particular choice of the
off--shell momenta and the choice of gauge. 
This dependence is not important since the final physical results will not
depend on this momentum choice, and will be gauge invariant.
We always choose the sum of all external momenta
to be equal to zero. In the case of the four quark operators
all four external lines carry momenta that are equal up to an 
overall sign.

The renormalization coefficient of the fermion field, $Z_q^{1/2}$,
is defined from the full propagator,

\begin{equation}\label{Zqdef}
    Z_q(\mu;a) = -{i\over 48}\mbox{Tr}\left(
      \gamma^\mu{\partial S_{bare}(p;a)^{-1}
        \over\partial{p_\mu}}\right)_{p^2=\mu^2}
\end{equation}
and can be calculated numerically using the Ward Identity
for the conserved vector current  $V^C$ in the form

\begin{equation}\label{WIvector}
    \Lambda^\mu_{V^C}(p;a) = -i{\partial 
         S_{bare}(p;a)^{-1}\over \partial p_\mu} \;,
\end{equation}
which implies

\begin{equation}\label{Zpsi}
  Z_q(\mu;a) = {1\over 48} \mbox{Tr}\left(\gamma^\mu 
                \Lambda^\mu_{V^C}(p;a)\right)_{p^2=\mu^2} \;.
\end{equation}
Substituting this into Eq.~(\ref{rencond}) for $V^C$ one can see that
the RI renormalization condition for the conserved vector current
is consistent with the equality $Z_{V^C}=1$.

\section{Scaling technique}

\label{sec:Scaling}

Our scaling technique exploits the renormalizability of lattice
QCD which implies that, when scaling violations are small,
the values of a renormalization coefficient calculated at
two different values of the lattice spacing will be related to each
other by an overall multiplicative coefficient independent of
the physical conditions such as physical volume or momentum scale,

\begin {equation}\label{rescaling}
  {Z_O(\mu, V; a')\over Z_O(\mu, V; a)} = R_O(a', a)
\end{equation}
We use this property to ``sew together'' evolutions of renormalization
coefficients over overlapping regions of momentum scale. We are going to 
describe this technique in detail here.

In the formula above we explicitly show the dependence of the 
renormalization
coefficients 
on the physical volume $V$.
While this dependence could be eliminated by taking the infinite volume
limit, in our scaling method this is not necessary for the intermediate
steps where the finite volume effects are explicitly taken into account.
To avoid complications associated with the quark mass renormalization,
the chiral limit is always taken.

In any Monte--Carlo QCD simulation the size of the lattice in any
direction is finite and rarely exceeds 32. This puts limits on the
momenta accessible in a simulation with one fixed value of the coupling
constant $\beta$. The components of momenta can assume only a discrete
set of values,
\begin{equation}\label{momenta}
  p_i a = {2 \pi \over L_i} n_i,   \ \ \ \ \ \
    n_i = -{(L_i-1)/2}, ..., 0, ..., {L_i/2} \;.
\end{equation}
For the momenta with large values of $n_i$ the discretization errors
become significant.
This forces one to use momenta
with small $n_i$ and therefore limits the range over which
the non--perturbative renormalization technique outlined above can be
carried out. 

Finding a way to study the renormalization of operators in a wide range
of momentum scales is important in many applications.
One particular but very important example is the connection with 
perturbation theory. Most of the lattice simulations are
performed at low momentum scales, where the non--perturbative
effects play a major role. On the other hand, 
if one needs to relate the results
to perturbation theory, for example, to use the Wilson coefficients
determined from a perturbative calculation,
one has to have access to large physical momenta
(several GeV) where 
perturbation theory should become accurate.
This is very difficult to
achieve in a simulation at a single value of $\beta$.

\begin{figure}[htb]
\epsfxsize=150mm
\epsfbox{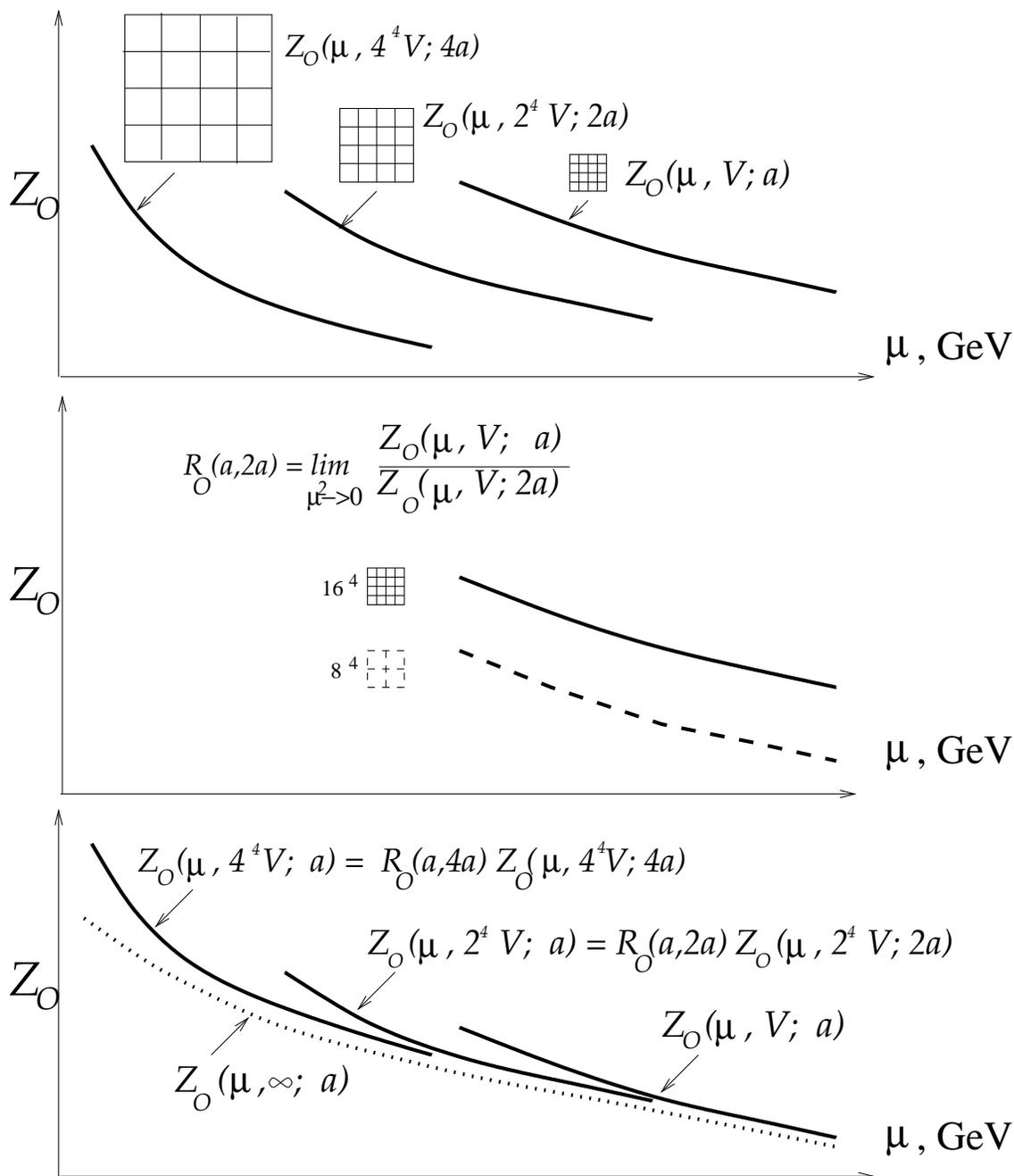}
\caption{Scaling procedure. Panel 1 shows $Z$--factors obtained at
a series of different $\beta$'s and same size lattices
corresponding to different physical volumes. These $Z$--factors
cannot be meaningfully compared directly since they are defined
with different values of lattice cutoff. Panel 2 shows the procedure
of obtaining the rescaling coefficients $R_O$. Panel 3 shows
$Z$--factors rescaled using $R_O$ and therefore defined with the
same lattice cutoff $a$. The ratios of the $Z$--factors on this
graph have direct physical meaning and are cutoff--independent.
Since different physical volumes are used, these $Z$--factors
do not necessarily overlap at the corresponding momenta.}
\label{fig:scalingprc}
\end{figure}

It is natural to think of using the results of numerical 
simulations at several $\beta$'s with overlapping regions
of physical momenta with small discretization errors 
to extend the range over 
which renormalization coefficients are
calculated numerically.
The procedure implementing this idea is the following (see Figure 
\ref{fig:scalingprc}).
One starts at $\beta(a)$ which is sufficiently large to insure that
there are momenta well below the lattice cutoff that are in
the perturbative regime. 
At such scales the renormalization coefficients
obtained from the Monte--Carlo simulation can be reliably related to
the perturbative ones. The finite volume effects at this $\beta(a)$
can be eliminated by numerically taking the infinite volume limit.
Alternatively, one can use perturbation theory carried out at finite volume.

The next step involves simulations at a series of $\beta$'s with
lattice spacing $a$ increasing by factors of two. The number of lattice
sites in all directions is fixed. The physical volume is finite and
different in each simulation, increasing by factors of $2^4$.
The range of momenta with small discretization errors in each simulation
also scales by factors of two. The overall range of momentum scales
with small discretization errors covered in this sequence of
 simulations can be very large.
But the $Z$--factors obtained with
different lattice spacings cannot be meaningfully compared directly,
 since the lattice spacing plays
the role of the ultraviolet cutoff on the lattice.
Changing the cutoff requires complex, generally nonperturbative
renormalization of operators in order to keep the physical,
low energy predictions of the theory unchanged.
Therefore, in order to relate the $Z$--factors obtained in 
simulations at different
$\beta$'s, it is necessary to rescale them
to be defined with the same ultraviolet cutoff equal to
the lattice spacing $a$ at some fixed value of $\beta$.
To achieve this, we use the property Eq.~(\ref{rescaling}) which
lies in the heart of our scaling method.
For the common ultraviolet cutoff we choose, quite arbitrarily,
the lattice spacing at $\beta=7.05$ from the first simulation in the series.

In order to accomplish the program of redefining all of the
$Z$--factors at a common ultraviolet cutoff $a$ it is necessary
to know the coefficients $R_O(a,a')$ that relate the $Z$--factors
computed at the cutoff $a'$ to the $Z$--factors that would be found
if the cutoff $a$ was used instead.
To compute the rescaling coefficients $R_O(a,a')$
we perform a series of additional simulations. Each additional
simulation has the lattice spacing two times larger than
the corresponding simulation from the main series. 
At the same time, the number of the 
lattice sites in each direction is reduced by a factor of two.
With this choice of parameters, the physical volume and the
values of physical momenta are the same for these two simulations.
If discretization errors were negligible, the $Z$--factors
from these two simulations computed at the same physical momenta
would be related by a momentum--independent
factor $R_O(2^na,2^{n+1}a)$, in accordance with Eq.~(\ref{rescaling}).
Here $2^na$ and $2^{n+1}a$ are the lattice spacings used correspondingly
in the simulation from the main series and the additional simulation,
while $a$ is the value of the common ultraviolet cutoff chosen to be
equal to the
lattice spacing used in the simulation with the largest $\beta$ in the
series.
 To find this factor we would use
the ratio $Z(\mu,V;2^na)/Z(\mu,V;2^{n+1}a)$ calculated at any value of physical
momentum scale $\mu$. Here $V$ is the physical volume in both simulations.

When discretization errors are present, $Z$--factors, as well as their
ratios, have momentum--dependent corrections. For the domain wall fermion
action these errors are quadratic in momentum,
\begin{equation}\label{ratioaOp2}
 {Z_O(\mu,V;2^na)\over{Z_O(\mu,V;2^{n+1}a)}} = 
     R_O(2^na,2^{n+1}a) + O\left((\mu 2^n a)^2\right) \;.
\end{equation}
We use quadratic in $\mu a$ fits to remove the $O\left((\mu 2^n a)^2\right)$
term and extract $R_O(2^na,2^{n+1}a)$.
By combining the rescaling coefficients obtained for pairs of
lattice spacings, $R_O(a,2a)$, $R_O(2a, 4a)$, etc., we can compute
the rescaling coefficients $R_O(a,2^na)$,
\begin {eqnarray}
\nonumber
  R_O(a,2a)\\
\label{R_O}
  R_O(a,4a)&=&R_O(a,2a)R_O(2a,4a)\\
\nonumber
  R_O(a,8a)&=&R_O(a,2a)R_O(2a,4a)R_O(4a,8a)\\
\nonumber \ldots\;,
\end{eqnarray}
or, more generally, $R_O(2^na,2^ma)$ where $n$ and $m$ are integers.

With the rescaling coefficients on the left hand side of equations
Eq.~(\ref{R_O}) we finally can apply the same formula 
Eq.~(\ref{rescaling}) to
the $Z$--factors computed in the main series of simulations at
different ultraviolet cutoffs to compensate for the non--trivial
renormalization required when different values of the cutoff are used.
In particular, from the original $Z$--factors in the main series,
$Z_O(\mu,V;a)$, $Z_O(\mu,2^4V;2a)$, $Z_O(\mu,4^4V;4a)$, etc.,
we obtain $Z_O(\mu,V;a)$, $Z_O(\mu,2^4V;a)$, $Z_O(\mu,4^4V;a)$, etc.,
\begin {eqnarray}
\nonumber
  Z_O(\mu,V;a)\\
\nonumber
  Z_O(\mu,2^4V;a)&=&R_O(a,2a) Z_O(\mu,2^4V;2a)\\
\label{rescale}
  Z_O(\mu,4^4V;a)&=&R_O(a,4a) Z_O(\mu,2^4V;4a)\\
\nonumber \ldots \;.
\end{eqnarray}
After such a rescaling the ratios of the new
$Z$--factors from all of the simulations are completely 
physical and represent the dependence
of the corresponding quantities on the momentum scale
and volume.
We would like to emphasize that direct computation of the $Z$--factors
on the left hand side of Eq.~(\ref{rescale}) in the whole
momentum scale range for which they have been determined would require
tremendous computer resources and at the very least would be completely
impractical.

After the rescaling, the renormalization coefficients $Z_O$ 
from different simulations evaluated at the same physical momentum scale will
in general differ, because they are calculated in different
physical volumes. In fact, this method offers one the possibility
to study the finite volume effects using same size lattices with
nevertheless quite different physical volumes, which can
provide certain advantages. The difference caused by the finite
volume effects gets smaller when the momentum scale becomes large
in comparison with the inverse size of the volume extent.
For some renormalization
coefficients these effects are quite large (see results for
the scalar and pseudoscalar densities and $O_{LL}^{\Delta{S}=2}$).
This effect is illustrated schematically in Figure \ref{fig:scalingprc}
on the graph at the bottom. The labels on the lines show explicit
dependence on the physical volume, which is different for the simulations
at different $\beta$'s. If the finite volume effects are small, these
lines overlap at the corresponding momentum scales. This is (almost)
the case for the local and exactly conserved vector and axial currents.
For any renormalization coefficient there exists a line that corresponds
to the infinite volume. It is shown as a dotted line on the plot
at the bottom of Figure \ref{fig:scalingprc}. 
Since in our method we do not need to take
this limit, our lines are just approximations of the infinite volume
line. At each momentum scale the line that has larger physical volume
is a closer approximation.

Another reason why the $Z$--factors computed at the same physical
momentum can differ between simulations and in fact from their true
values is the presence of discretization
errors. We eliminate these errors from the determination of rescaling
coefficients $R_O$ using the procedure outlined above, but not from the
$Z$--factors themselves. To remove discretization errors from
the $Z$--factors we could use even more additional simulations
with matching physical volumes so that a continuum limit at
each momentum could be taken. The presence of these errors does
not effect our ability to relate $Z$--factors between arbitrarily
different momenta since we only need $R_O$ for that.
Since discretization errors are quadratically small for small lattice
momenta, the values of the computed $Z$--factors at each physical momentum
are closer to their true value for data points that come from the simulations
at larger $\beta$. When the finite volume effects are negligible, so that
the expected difference between $Z$--factors at the corresponding physical
momenta is small, one can use this observation to estimate the
magnitude of the discretization error effects. It turns out
that their significance varies depending on the particular operator
for which the $Z$--factor is computed. As a result of our study
we found that discretization errors are quite large
for the local vector and axial currents, and even larger for the
exactly conserved vector and axial currents. In the $16^4$ volume,
for example, only four to five points with the lowest lattice
momenta have reasonably small errors. This implies that it is virtually
impossible to study the running of the renormalization coefficients
for these operators without using the scaling technique or performing
simulations in very large volumes that is extremely expensive
computationally.

\section{Numerical Results}

\label{sec:Results}

We use the fermion action which is the hermitian conjugate of the action in 
Ref.~\cite{Shamir:1993zy} where the domain wall fermion method was
introduced. The domain wall fermion operator that we use is given by
\begin{equation}\label{DWFoperator}
D_{x,s;x',s'}=\delta_{s,s'}D^{||}_{x,x'}+\delta_{x,x'}D^{\bot}_{s,s'}
\end{equation}
\begin{equation}
D^{||}_{x,x'}={1\over 2}
\sum_\mu\left[(1-\gamma_\mu)U_{x,\mu}\delta_{x+\hat{\mu},x'}
+(1+\gamma_\mu)U^\dag_{x',\mu}\delta_{x-\hat{\mu},x'}\right] 
+ (M_5-4)\delta_{x,x'}
\end{equation}
\begin{eqnarray}
\nonumber
D^\bot_{s,s'} &=& {1\over 2}\left[(1-\gamma_5)\delta_{s+1,s'}
+(1+\gamma_5)\delta_{s-1,s'}-2\delta_{s,s'}\right]\\
&&-{m_f\over 2}\left[(1-\gamma_5)\delta_{s,L_s-1}\delta_{0,s'}
+(1+\gamma_5)\delta_{s,0}\delta_{L_s-1,s'}
\right]
\end{eqnarray}
Operators are constructed from an interpolating operator for the light
fermion of the form
\begin{eqnarray}
  q(x) &=& P_L \Psi_{x,0} + P_R \Psi_{x,L_s-1}\\
  \bar{q}(x) &=& \overline{\Psi}_{x,0} P_R + \overline{\Psi}_{x,L_s-1} P_L
\end{eqnarray}
Evaluation of the matrix elements is done in Landau gauge. The gauge
is fixed on each configuration by finding a unitary transformation of 
the lattice links
that maximizes the functional

\begin{equation}
  \mbox{Tr} \sum^4_{\mu=1} \left( U_\mu(x) + U_\mu^\dag(x) \right) \;.
\end{equation}

In our calculations the parameter $M_5$ is set to $1.8$.
We use lattice volumes $8^4$ and $16^4$
at three values of $\beta$, 6.0, 6.45,
and $7.05$. We assume the inverse lattice spacing at $\beta = 6.0$ to be
$1.96 GeV$, twice that at $\beta=6.45$, and four times that
at $\beta=7.05$. These values are based on the string 
tension data in \cite{Bali:1993ru}. The scale for $\beta=6.45$
was obtained by interpolating the data from that paper using the
two--loop perturbation formula for the running coupling constant.
 The result for $\beta=7.05$ was obtained
by extrapolating and is less accurate.

The chiral limit for the local operators
is taken using correlated fits to the results
obtained at a set of masses $m_f=0.004, 0.012, 0.020$
in $16^4$ volume and $m_f=0.004, 0.012, 0.020, 0.028$ in $8^4$ volume.

In the case of conserved currents all simulations are performed with
$m_f=0.05$, which is small enough for the results to be in the chiral
limit. We tested this statement by repeating one of the points
using $m_f=0.02$ and found the data to be in agreement
 within statistical errors.

\subsection{Local Vector and Axial Currents}

\label{sec:local}

Now we present the results for renormalization coefficients
of local vector and axial currents.
The local vector and axial current densities are defined by
\begin{equation}
   V_L^\mu (x) = \bar{q}\gamma^\mu{q} (x)
\end{equation}
and
\begin{equation}
   A_L^{\mu} (x) = \bar{q}\gamma^\mu\gamma_5{q}(x) \;.
\end{equation}
In these equations flavor non--singlet currents are implied. 
The flavor index is suppressed to simplify the notation.
The matrix elements of the vector and axial currents 
in external quark states include finite terms
that depend on the direction of the external momentum
in addition to the term proportional to the free--field vertex.
In the continuum the form of these matrix elements is the following,

\begin{equation}\label{vecform}
    \Lambda_{V^L}^\mu(p) = A_{V^L}(p^2) \gamma^\mu + 
           B_{V^L}(p^2) {p^\mu\pslash{p}\over{p^2}}
\end{equation}
and
\begin{equation}\label{axform}
    \Lambda_{A^L}^\mu(p) = A_{A^L}(p^2) \gamma^\mu\gamma_5 + 
           B_{A^L}(p^2) {p^\mu\pslash{p}\over{p^2}}\gamma_5 \;.
\end{equation}
For the $Z$--factors we use the definition with the four Lorentz components
of matrix elements combined by contracting with the gamma matrices,
\begin{equation}
  {Z_q(\mu;a)\over Z_{V^L}(\mu;a)} \equiv
        {1\over 48}
  \mbox{Tr}\left(\Lambda_{V^L}^\mu(p;a)\gamma^\mu\right)_{p^2=\mu^2}
\end{equation}
and
\begin{equation}\label{ZAlocal}
  {Z_q(\mu;a)\over Z_{A^L}(\mu;a)} \equiv
        {1\over 48}
\mbox{Tr}\left(\Lambda_{A^L}^\mu(p;a)\gamma^\mu\gamma_5\right)_{p^2=\mu^2} \;.
\end{equation}
With these definitions the second term in 
Eqs.~(\ref{vecform}) and (\ref{axform}) with coefficient $B(p^2)$ 
is included as part of the matrix element. (Alternatively, one could
separate the coefficient of the term proportional to the free--field 
vertex, for example, by tracing the $\mu=0$ component of the
matrix element calculated at momenta $p$ with $p^0=0$ with $\gamma^0$
for the vector current and $\gamma^0\gamma_5$ for the axial current.)

\begin{figure}
\epsfxsize=15cm
\epsfbox{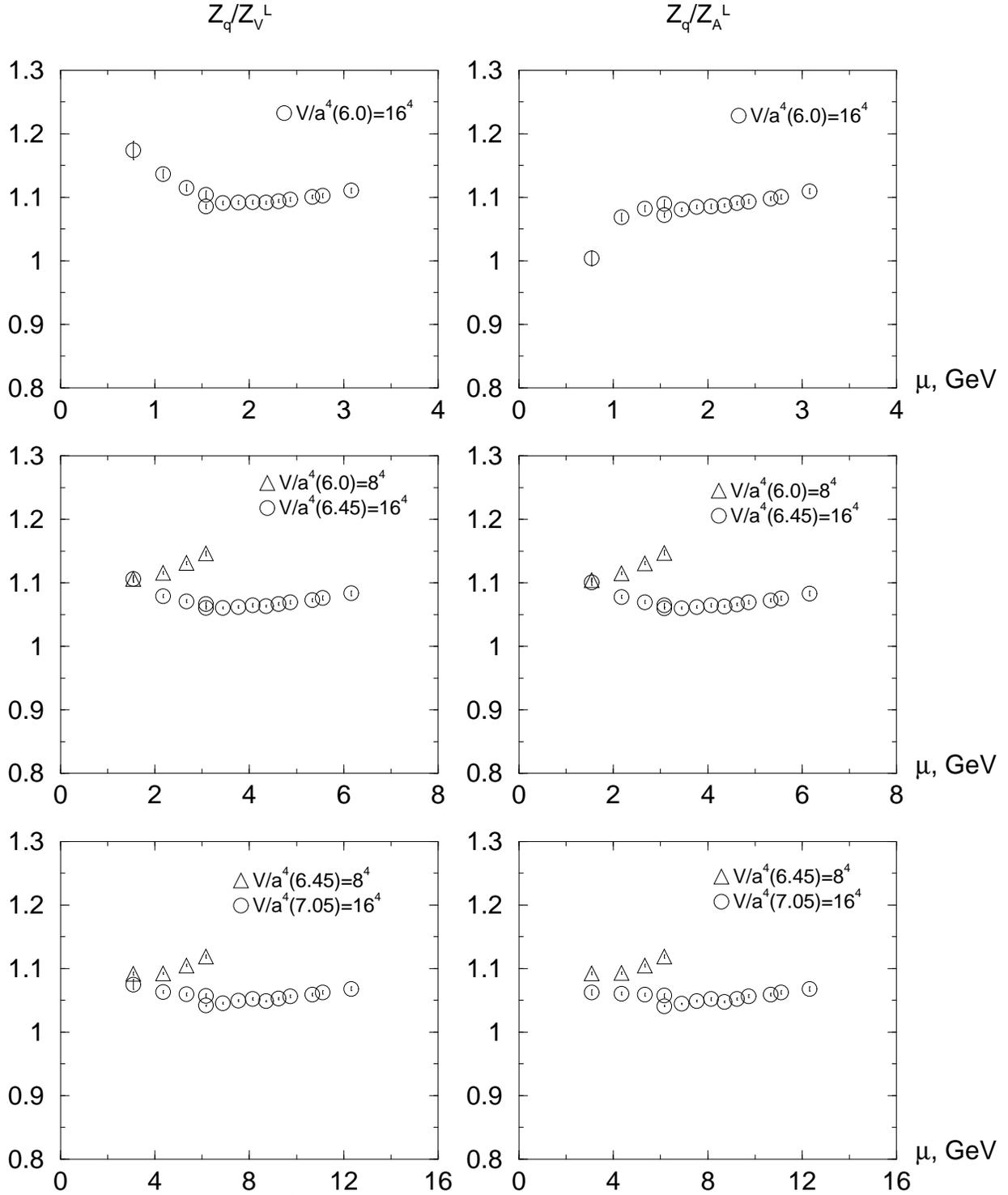}
\caption{$Z_q/Z_{V^L}$ and $Z_q/Z_{V^L}$ in the chiral limit
from local vector and axial currents. Each pair of graphs is for the
same physical volume.}
\label{fig:ZvZa}
\end{figure}

Figure \ref{fig:ZvZa} shows the renormalization
coefficients $Z_q/Z_{V^L}$ and $Z_q/Z_{A^L}$ versus
the energy scale $\mu$ in GeV  in the chiral limit.
The rows correspond to different physical volumes.
The pairs of points on each plot in the two lower rows are from
simulations at pairs of different $\beta$'s with the lattice spacing
differing by factors of two. Since the number of lattice sites differs
by the inverse of two, the physical volumes are the same.

\begin{figure}
\epsfxsize=15cm
\epsfbox{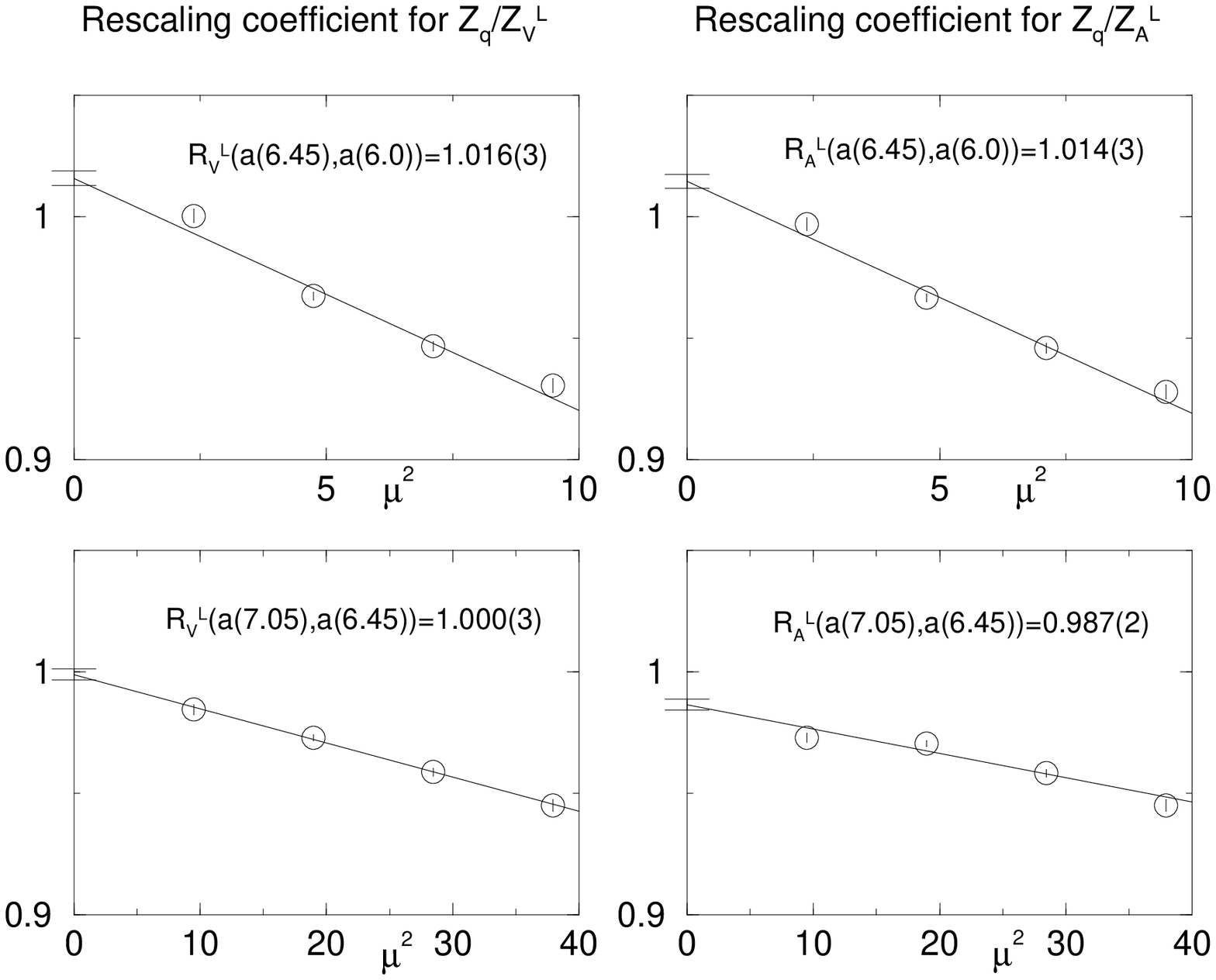}
\caption{The ratios of the renormalization factors $Z_q/Z_{V^L}$ and
$Z_q/Z_{A^L}$ obtained
at different $\beta$'s in the same physical volumes versus momentum
squared. The deviation from constant values are due to 
order $(\mu{a})^2$ discretization errors.
Linear fits in $\mu^2$ are used to find the rescaling coefficients
between renormalization factors at different $\beta$'s.}
\label{fig:ZvZascalefit}
\end{figure}

Figure \ref{fig:ZvZascalefit} demonstrates the procedure of 
obtaining the rescaling coefficients $R_A$ and $R_V$ for two pairs
of $\beta$, $6.45,6.0$ and $7.05,6.45$. Each point in 
Figure \ref{fig:ZvZascalefit}
is a ratio of the $Z$--factors in Figure \ref{fig:ZvZa} determined
at the same physical momentum and in the same physical volume.
There are four ratios at each pair of $\beta$'s, since we only use
the lowest four momenta in $8^4$ lattice volume
(triangular symbols in Figure \ref{fig:ZvZa}).

As described in Section \ref{sec:Scaling}, these ratios would be
$\mu$ independent in the absence of discretization errors. With finite
lattice spacing these errors are quadratic in momentum scale $\mu$,
see Eq.~(\ref{ratioaOp2}). 
In order to remove the effects of discretization
errors, we extrapolate the ratios linearly in $\mu^2$ to $\mu\rightarrow{0}$,
Figure \ref{fig:ZvZascalefit}.
 The rescaling coefficients for the local vector and axial density operators
turn out to be very close to one.
 
\begin{figure}
\epsfxsize=15cm
\epsfbox{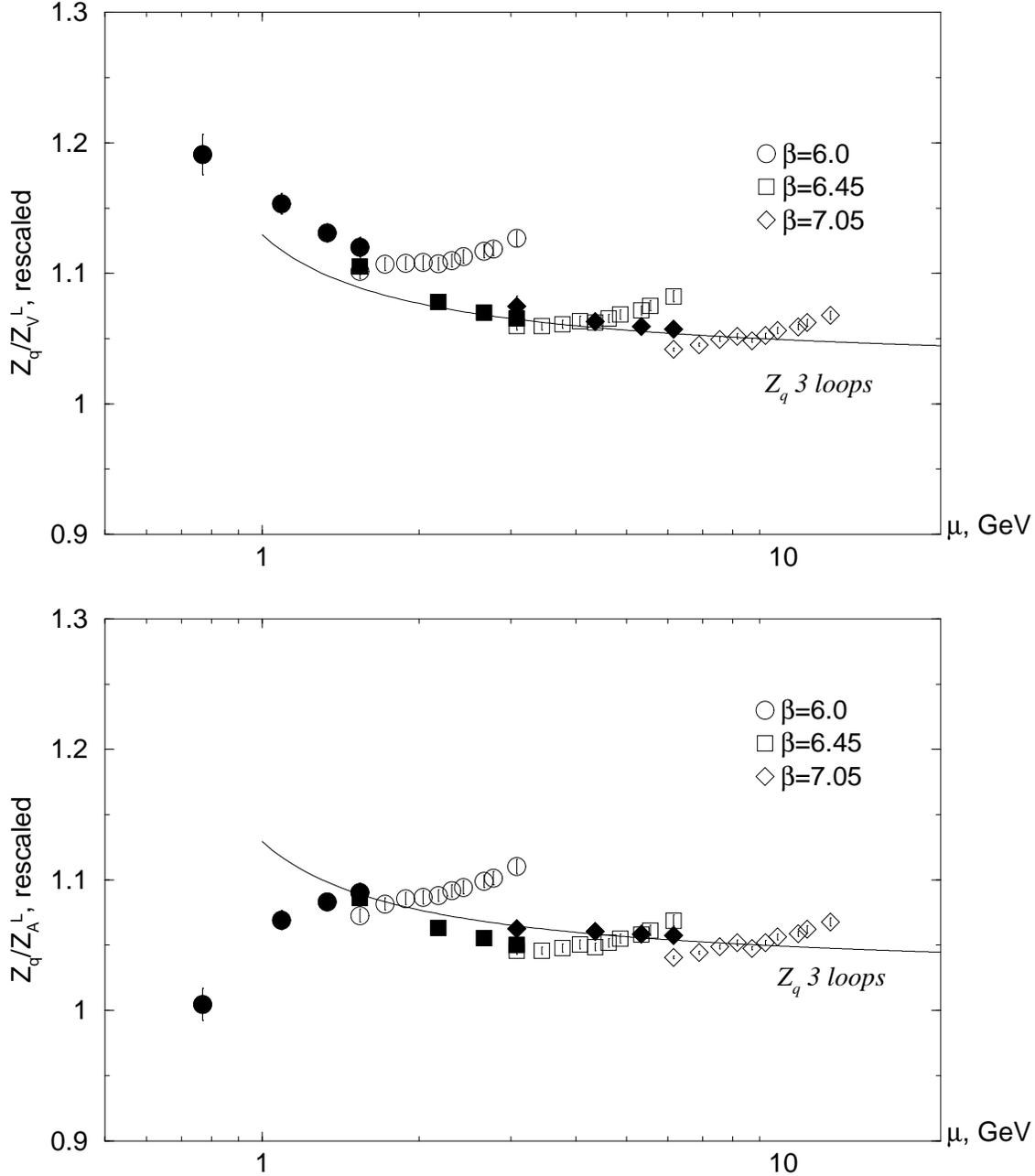}
\caption{$Z_q/Z_{V^L}$ and $Z_q/Z_{A^L}$ 
for local vector and axial currents in a wide range of scales $\mu$. 
All graphs have the same ultraviolet cutoff equal to
the lattice spacing at $\beta=7.05$. 
Different symbols correspond
to data in different physical volumes (labeled by $\beta$ at which they are
obtained). Solid symbols are for data with small discretization errors.}
\label{fig:ZvZaAllbeta}
\end{figure}

Finally, using the rescaling coefficients we rescale the 
renormalization coefficients according to Eq.~(\ref{rescale})
so that they are all defined with the same ultraviolet cutoff, determined by
the lattice spacing at $\beta=7.05$. The results are shown in Figure 
\ref{fig:ZvZaAllbeta}. There is a pretty good agreement between
the data computed in different physical volumes at low lattice momenta
(the first few low momentum points on each graph) that indicates
small finite volume effects for local vector and axial current data.
At the same time, discretization errors start effecting data pretty
early. To emphasize the actual behavior
of the $Z$--factors we use solid symbols
for the points that have relatively small discretization errors.
These are the same points that are used in calculating the rescaling
coefficients in Figure \ref{fig:ZvZascalefit}.

The renormalization
group running of $Z_q$ from perturbative analysis \cite{Chetyrkin:2000pq}
is represented by a solid line. This line is defined up to an overall 
multiplicative constant that we do not determine. Instead, we choose
it so that the line agrees with our numerical data at scales about 10~GeV.
The coefficient $Z_{V^L}$ should be momentum--independent.
It reflects the renormalization of the local vector current
relative to the exactly conserved on the lattice vector current
whose renormalization coefficient is equal to one according to
the Ward Identity, Section \ref{subsec:Conserved}.
The two currents are different at the scales of order $a^{-1}$,
while for the scales $\mu\ll a^{-1}$ the difference 
can be absorbed into $Z_{V^L}$.
 Therefore,
the dependence of $Z_q/Z_{V^L}$ on the scale $\mu$ should be the
same as the dependence of $Z_q$ itself. 
There is a good agreement between the RG line
and the data for  $Z_q/Z_{V^L}$ in the whole range of the scales.
On the other hand, $Z_{A^L}$ defined by Eq.~(\ref{ZAlocal}) is
momentum--independent only at large scales $\mu$. The reason is
the same as for the difference between $Z_q$ obtained from the
conserved vector current and $Z_q^{(A)}$ from axial current, see
Section \ref{subsec:Conserved}.

\subsection{Scalar and Pseudoscalar operators}

Now we present the results for renormalization coefficients of scalar
and pseudoscalar operators.
The matrix elements of the pseudoscalar density at low energy scales (about
1--2~GeV) receive contributions from the chiral 
symmetry breaking in the form of a pole in momentum squared and quark mass,

\begin{equation}\label{PseudoPole0}
   \Lambda_P(p^2) = {Z_q(p^2)\over Z_P(p^2)} + {C\over p^2}
        {\langle \bar{q}q\rangle\over m_f} \;.
\end{equation}
The general form of the chiral condensate is given by the equation
\begin{equation}\label{pbp}
\langle \bar{q}q\rangle = a + b m_f + {c\over m_f \sqrt{V}}
\end{equation}
The third term in the above equation is due to the 
fermion zero modes present in our
quenched simulations, Ref.~\cite{Wingate:2000jx}. This term, which
is inversely proportional to $m_f$, is
suppressed by the square root of the 
 volume and is small in comparison with the constant term
for the parameters used in our simulations.
 Therefore, in Eq.~(\ref{PseudoPole0}) for the pseudoscalar density
the mass dependence of the second term is very well described by
the equation
\begin{equation}\label{PseudoPole}
{C_1\over p^2 m_f} \;.
\end{equation}
For the scalar density we have,
\begin{equation}\label{ScalarPole0}
   \Lambda_S(p^2) = {Z_q(p^2)\over Z_S(p^2)} + {C\over p^2}
        {\partial\langle\bar{q}{q}\rangle\over\partial m_f} \;.
\end{equation}
In this case the constant term in the chiral condensate,
Eq.~(\ref{pbp}),
is removed by the derivative. Therefore, 
the third term in Eq.~(\ref{pbp}) which is inversely proportional to $m_f$,
can determine the form of the mass--dependence of $\Lambda_S$
in small physical volumes. In this case the contribution 
to $\Lambda_S(p^2)$ from the second term in Eq.~(\ref{ScalarPole0})
has the form
\begin{equation}\label{ScalarPole} 
  -{1\over m_f^2}{C_2\over p^2} \;.
\end{equation}
When taking the chiral limit of scalar and pseudoscalar densities
at energy scales of a few GeV in finite volume, one needs to remove
these non--leading, $1/p^2$ contributions which have singular
behavior as $m_f\rightarrow 0$.
For a thorough discussion of both of these effects see \cite{Chris:2000}.

\begin{figure}
\epsfxsize=15cm
\epsfbox{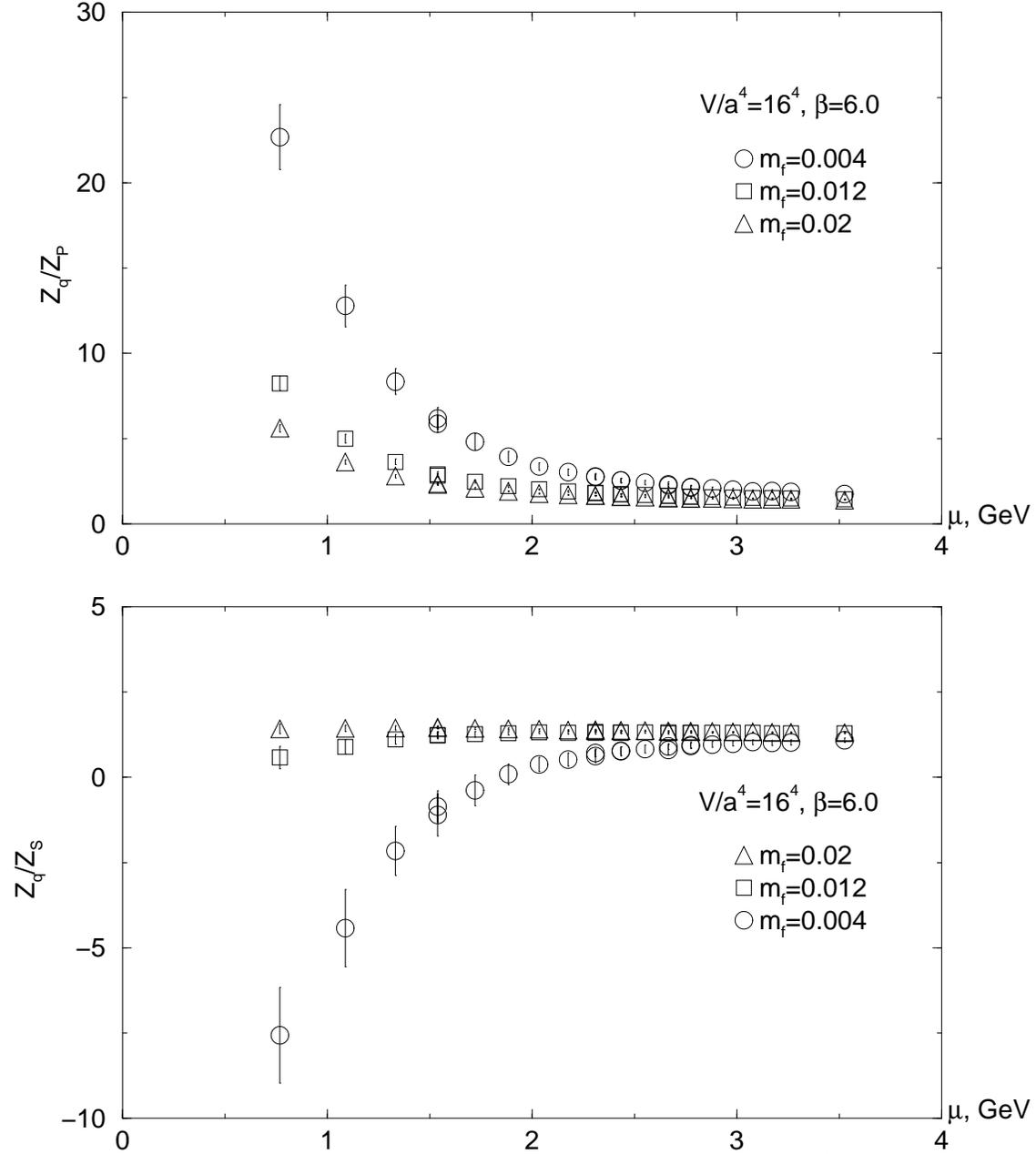}
\caption{Mass poles in the pseudoscalar and scalar operators
in a $16^4$ lattice volume at $\beta=6.0$. These poles have 
to be subtracted at each momentum.}
\label{fig:ZpZsPole}
\end{figure}

\begin{figure}
\epsfxsize=15cm
\epsfbox{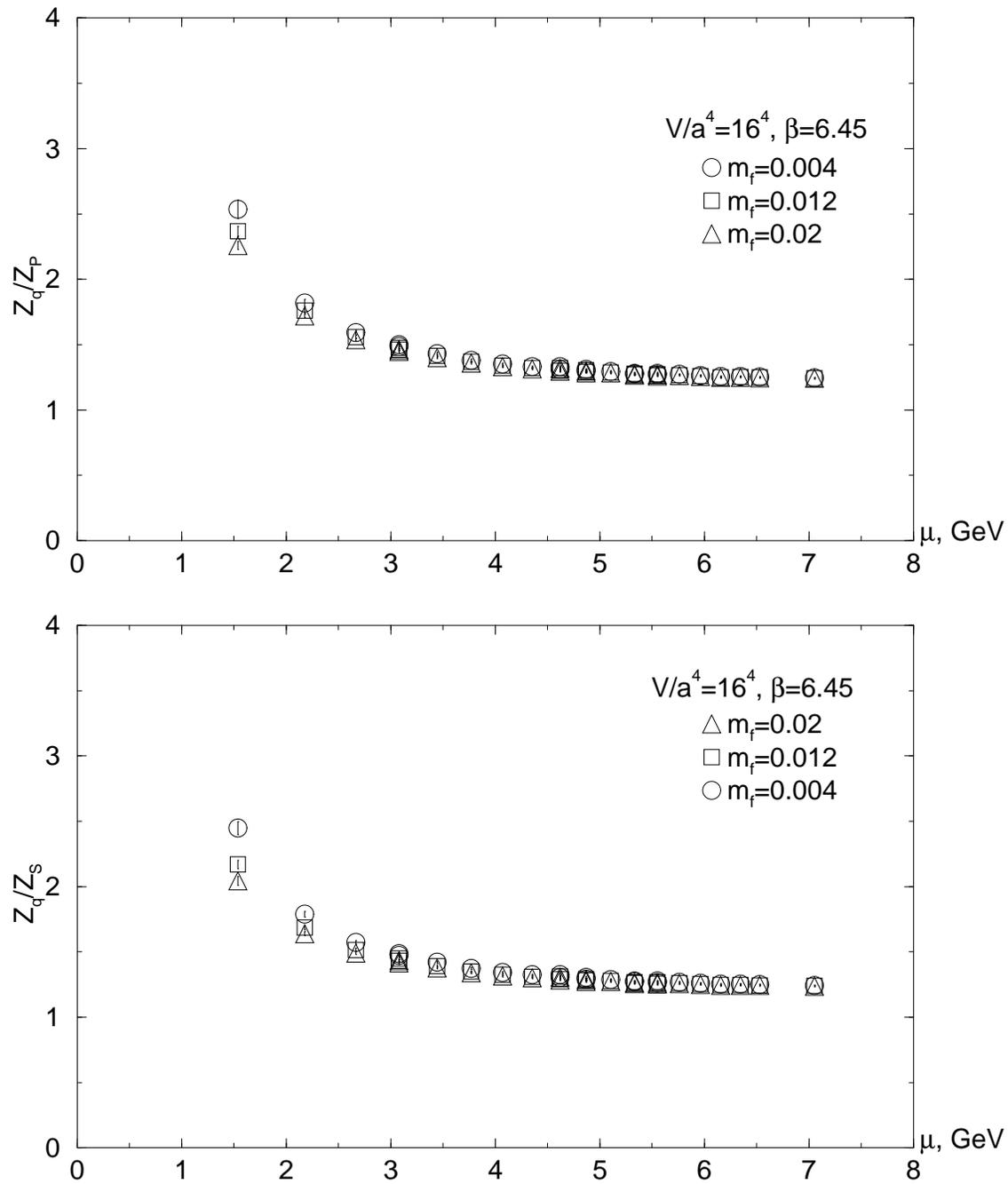}
\caption{Mass pole effects are much smaller at $\beta=6.45$ in 
 $16^4$ lattice volume.}
\label{fig:ZpZsPole645}
\end{figure}

The presence of these terms is clearly seen
in Figure \ref{fig:ZpZsPole} where the
data for a $16^4$ volume at $\beta=6.0$ 
for a set of quark masses is presented.
Their contribution to the scalar density becomes 
significant only for $m_f=0.004$ and is almost invisible for $m_f=0.02$.
Figure \ref{fig:ZpZsPole645} shows results at   $\beta=6.45$. At this
$\beta$ there is little evidence of such small--$m_f$ divergent term,
which is consistent with the absence of the chiral symmetry breaking
in this smaller physical volume.

\begin{figure}
\epsfxsize=15cm
\epsfbox{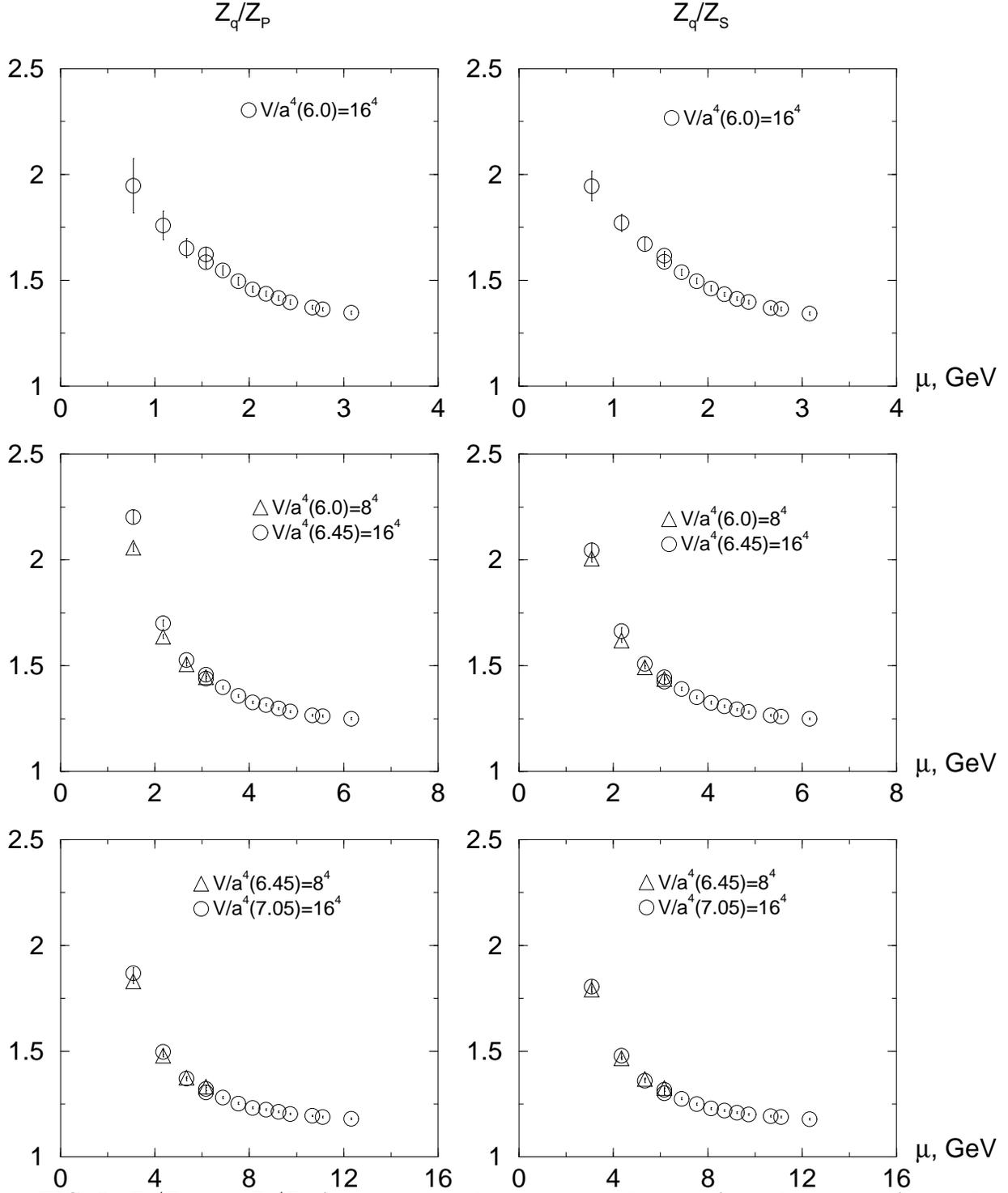}
\caption{$Z_q/Z_P$ and $Z_q/Z_S$ 
from pseudoscalar and scalar density after subtraction of $m_f$ poles.
The graphs from top to
bottom correspond to different physical volumes, while each
pair of graphs is for the same physical volume.}
\label{fig:ZpZs}
\end{figure}

Figure \ref{fig:ZpZs} shows the renormalization coefficients
$Z_q/Z_P$ and $Z_q/Z_S$ plotted versus the energy scale $\mu$
in GeV in the chiral limit. To obtain these
results, we performed correlated fits in $m_f$ at each momentum
in the form Eqs.~(\ref{PseudoPole}) and (\ref{ScalarPole}) plus finite, 
$m_f$--independent renormalization coefficients $Z_q/Z_P$ and $Z_q/Z_S$.
The arrangement of the graphs is the same as
for the local vector and axial currents.

\begin{figure}
\epsfxsize=15cm
\epsfbox{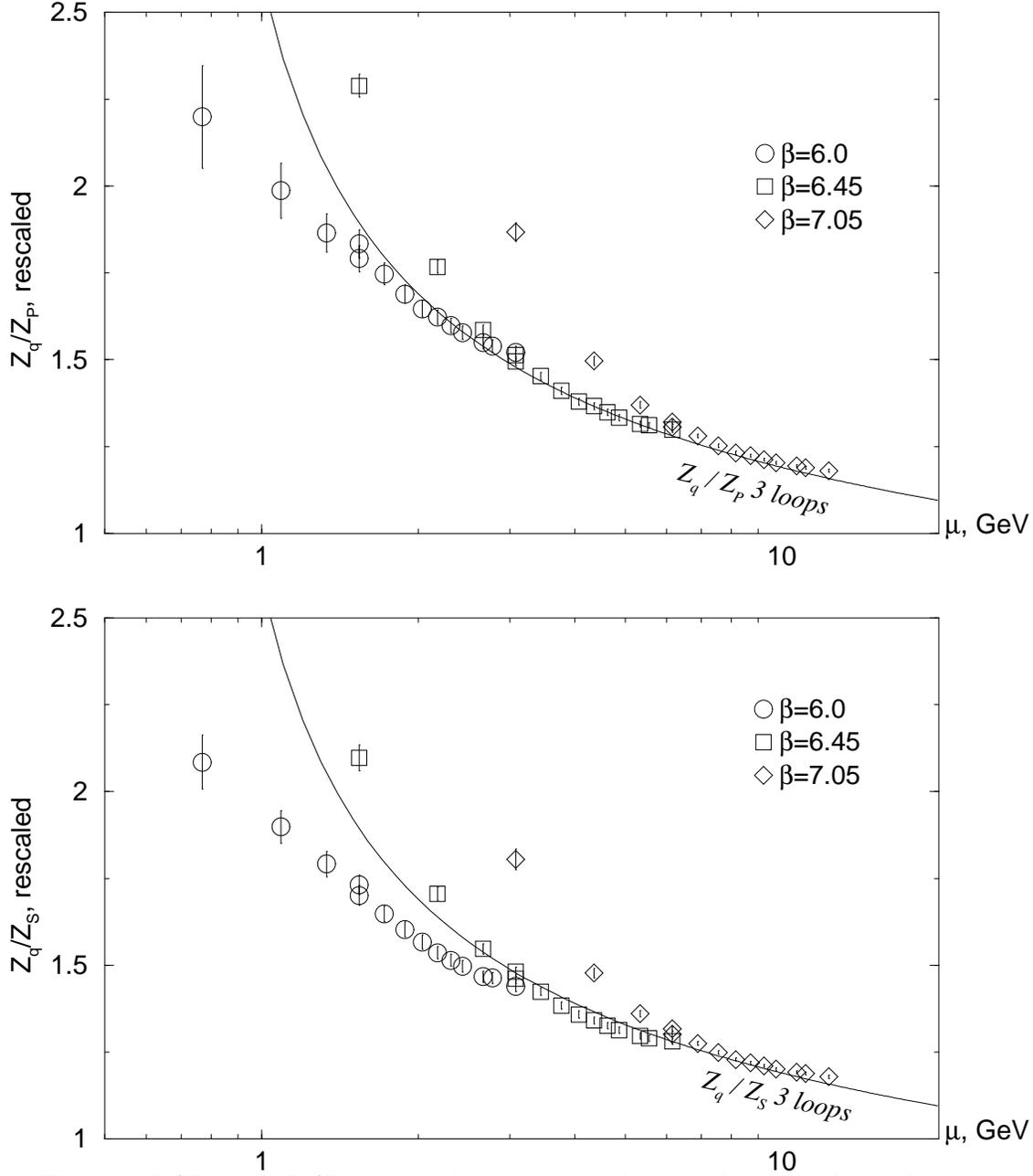}
\caption{$Z_q/Z_P$ and $Z_q/Z_S$ in a wide range of scales $\mu$.
All graphs have the same ultraviolet cutoff equal to
the lattice spacing at $\beta=7.05$. 
Different symbols correspond
to different physical volumes (labeled by the $\beta$ at which they are
obtained). The finite volume effects are
noticeable for momenta of the order of the inverse box size.}
\label{fig:ZpZsAllbeta}
\end{figure}

The result of the scaling procedure is shown in Figure
\ref{fig:ZpZsAllbeta}. Solid lines are the renormalization
group running from a $3$--loop perturbation theory analysis,
\cite{Chetyrkin:2000pq}. 
They represent
the evolution of the $Z$--factors with the energy scale, and
are defined up to an overall multiplicative constant
which we do not determine. We choose this
constant to match the RG curve with our data at large scales $\mu$
around $10$ GeV. The data exhibits large finite volume
effects at small momenta in each of the three different 
physical volumes. This results in the difference between
data at the corresponding momenta obtained in different volumes.
This difference vanishes 
as the momentum scale becomes large within the individual momentum
range for each volume.
On the other hand, discretization error effects are barely
noticeable.

\subsection{Conserved Vector and Axial Currents}

\label{subsec:Conserved}

The exactly conserved domain wall vector and axial currents can be used
to compute $Z_q$. Following Ref.~\cite{Furman:1995ky},
with our fermion operator given by Eq.~(\ref{DWFoperator}) 
these currents are defined by the equation

\begin{equation}\label{lbl:jmu}
  J_\mu(x) = \sigma_J \sum_{s=0}^{L_s/2-1} j_\mu(x,s)
         + \sum_{s=L_s/2}^{L_s-1} j_\mu(x,s) 
\;,
\end{equation}
where $J=V$ or $A$ and $\sigma_V=+1$, $\sigma_A=-1$ and
\begin{equation}\label{lbl:jmu1}
j_\mu(x,s) = 
\overline\Psi_{x+\hat\mu,s}\frac{1+\gamma_\mu}{2}U^\dagger_\mu(x)\Psi_{x,s}
- \overline\Psi_{x,s}\frac{1-\gamma_\mu}{2}U_\mu(x)\Psi_{x+\hat\mu,s}\;.
\end{equation}
Here flavor non--singlet currents are implied with the flavor indices
suppressed for simplicity of notation.
With the definition of $Z_q$ given by Eq.~(\ref{Zqdef}) we can use
the Ward Identity Eq.~(\ref{WIvector}) for the exactly conserved
vector current which implies Eq.~(\ref{Zpsi}). The corresponding 
Ward Identity for the axial current has an additional
term that does not vanish in the chiral limit, due to the
presence of a massless Goldstone boson. It becomes
negligible at sufficiently large scales $\mu$ (see a proof in
\cite{Martinelli:1995ty} for example). For this reason,
when we apply the same equation Eq.~(\ref{Zpsi}) with $\Lambda^\mu_{A^C}$,
the resulting renormalization coefficient $Z^{(A)}_q$ is
different from $Z_q$ when evaluated at scales below $\sim$2~GeV.

Due to the presence of the fermion fields at separate space--time 
points in Eqs.~(\ref{lbl:jmu}) and (\ref{lbl:jmu1}), propagators
 must be evaluated for sources at different points on the four--dimensional
lattice. In addition, 
the summation over the position in the fifth dimension implies that
propagators for all of these positions should in principle be
calculated. Therefore, the calculation of the matrix elements of
conserved currents can be very expensive. To minimize the amount of 
computational time, we use a random source estimator to 
compute the part of the sum
between $s=1$ and $s=L_s-2$, with propagators for $s=0$ and $s=L_s-1$
calculated explicitly. Also, instead of calculating all four components of
$\Gamma^\mu(p)$ for a given momentum $p$, we
calculate $\Gamma^0(p)$ for momenta related to $p$ by permuting
its $0$th component with the rest of the components,
\begin{eqnarray}\label{ccCombine}
\nonumber
  \mbox{Tr}\left(\gamma^\mu \Gamma^\mu(p^0,p^1,p^2,p^3)\right) &=&
   \mbox{Tr}\left(\gamma^0 \Gamma^0(p^0,p^1,p^2,p^3)\right)
   + \mbox{Tr}\left(\gamma^0 \Gamma^0(p^1,p^0,p^2,p^3)\right)\\
\label{trick}&&   + \mbox{Tr}\left(\gamma^0 \Gamma^0(p^2,p^1,p^0,p^3)\right)
   + \mbox{Tr}\left(\gamma^0 \Gamma^0(p^3,p^1,p^2,p^0)\right).
\end{eqnarray}
The computer time required to Fourier transform a matrix element is negligible
in comparison with the calculation of the matrix element itself, 
therefore the formula above allows us to obtain the result with 
only a quarter of the running time.

\begin{figure}
\epsfxsize=15cm
\epsfbox{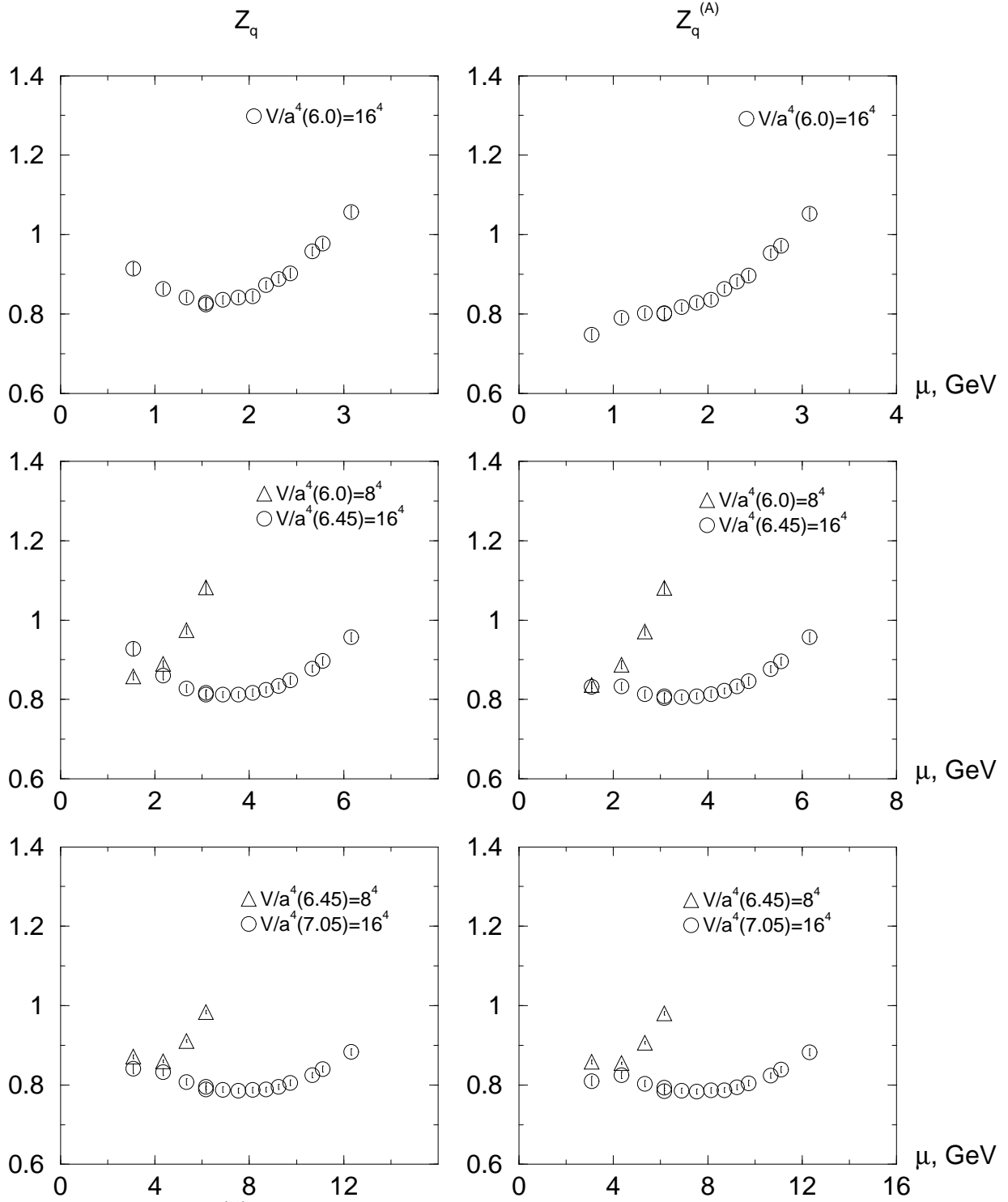}
\caption{$Z_q$ and $Z^{(A)}_q$
from conserved vector and axial currents. The graphs from top to
bottom correspond to different physical volumes, while each pair
of graphs is for the same physical volume.}
\label{fig:ZvcZac}
\end{figure}

We used $m_f=0.05$ in our simulations for the conserved vector and
axial currents which from our experience is sufficiently close to the 
chiral limit. A test run at $\beta=6.0$ with $m_f=0.02$ with
$30$ configurations produced results consistent within statistical errors.
In Figure \ref{fig:ZvcZac} we plot $Z_q$ computed from the conserved vector
current using Eqs.~(\ref{Zpsi}) and (\ref{ccCombine}). The same figure
shows the renormalization coefficient $Z^{(A)}_q$ obtained by applying these 
equations with the axial current.

\begin{figure}
\epsfxsize=15cm
\epsfbox{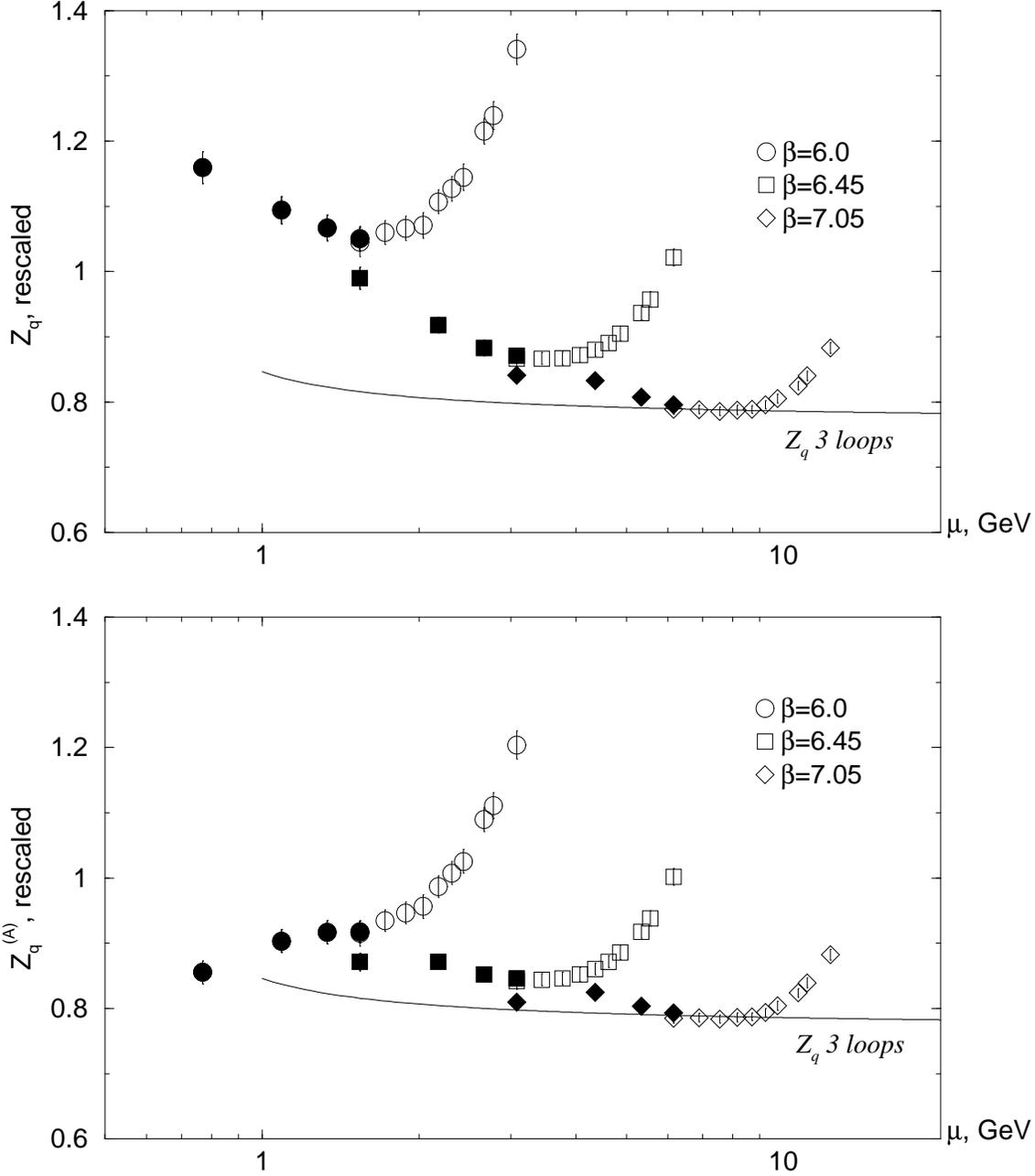}
\caption{$Z_q$ and $Z^{(A)}_q$ plotted for a wide range of scales $\mu$.
All graphs have the same ultraviolet cutoff equal to
the lattice spacing at $\beta=7.05$. 
Different symbols correspond
to data in different physical volumes (labeled by the $\beta$ at which they are
obtained). Solid symbols are for data with small discretization errors.}
\label{fig:ZvcZacAllbeta}
\end{figure}

The data is similar to that obtained for the local vector and axial 
currents, but shows larger discretization errors.
In Figure \ref{fig:ZvcZacAllbeta} we plot the evolution of $Z_q$ 
and $Z^{(A)}_q$ factors computed from the conserved
vector and axial currents after rescaling, such that the ultraviolet
cutoff is equal to the lattice spacing at $\beta=7.05$.
The renormalization
group running of $Z_q$ from a perturbative analysis \cite{Chetyrkin:2000pq}
is represented by a solid line. This line is defined up to an overall 
multiplicative constant that we do not determine. Instead, we choose
it so that the line agrees with our numerical data at scales about 10~GeV.

From the results in Figure \ref{fig:ZvcZacAllbeta} obtained
at different $\beta$'s, one can see that
while the finite volume effects are still small, as in the local
current case, the
quadratic in $\mu{a}$ deviations due to discretization errors 
that curves the results up become significant very early. Essentially,
only the first five points with lowest momenta at each $\beta$ seem
to have relatively small discretization errors. This statement of
course is correct only in a $16^4$ volume. But it does suggest that
attempts to analyze the momentum dependence of $Z_q$ from conserved
currents using simulations at a single value of $\beta$ are subject to
large discretization errors. We use solid symbols for the data points
 that have reasonable discretization errors.
These are the same points that are used in calculating the 
rescaling coefficients for conserved currents.

The results from the conserved vector and axial currents
agree pretty well at the scales of about 2~GeV and above.
At the same time, at 2~GeV and below there is large
difference between $Z_q$ and $Z^{(A)}_q$. The difference
reaches almost 20\% at about 1~GeV and is more than 30\% at 0.8~GeV.
The numerical results therefore indicate that the
additional finite term in the axial Ward Identity is large at these scales.
The same argument is consistent with the data for the rescaling coefficients
in Table \ref{table:1}. The agreement between rescaling coefficients for $Z_q$
and $Z^{(A)}_q$ is much better
for the pair of weaker couplings $\beta=6.45$ and $\beta=7.05$ than for
$\beta=6.0$ and $\beta=6.45$.

The results in Figure~\ref{fig:ZvcZacAllbeta}
are in poor agreement with the renormalization group curve
from perturbation theory, though it becomes
slightly better at large momenta as the scale dependence of 
$Z_q$ (and $Z^{(A)}_q$) becomes flatter. This is in sharp contrast
with the results for the local vector and axial currents,
Figure~\ref{fig:ZvZaAllbeta}. In the absence of scaling violations
the results from local currents should agree up to an overall 
momentum--independent factor with the results from the conserved currents,
as argued in Section~\ref{sec:local}.
We interpret this inconsistency between our results for the local
and conserved currents as arising from discretization errors present
in the non--local, conserved currents. These can arise in two ways.
First, our matching procedure using the small $8^4$ volumes might
not be able
to properly account for the order $O(a^2\mu^2)$ terms. This difficulty
can be addresses by working on $16^4$ and $32^4$ volumes
that would allow smaller lattice momenta to be used, therefore
reducing the errors. The second
source of errors may arise from the failure of these non--local
conserved currents to scale at couplings as strong as $\beta=6.0$.

\subsection{$B_K$}

\label{sec:bk}

In this section we present the results of non--perturbative renormalization
of a phenomenologically important parameter $B_K$ which is relevant to
the calculation of the $K^0-\bar{K}^0$ mixing amplitude. 
The standard definition of $B_K$ is given by
\begin{equation}
  B_K(\mu) = {\langle\bar{K}^0|{O}_{LL}^{\Delta{S}=2}(\mu)|K^0\rangle
  \over
\langle\bar{K}^0|{O}_{LL}^{\Delta{S}=2}(\mu)|K^0\rangle_{VSA}}
\label{bkdef}
={\langle\bar{K}^0|{O}_{LL}^{\Delta{S}=2}(\mu)|K^0\rangle
  \over
\frac{8}{3}f^2_K m^2_K} \;.
\end{equation}
This definition implies that $B_K$ renormalizes
in the same way as the four--fermion operator

\begin{equation}\label{OdS2}
  O_{LL}^{\Delta{S}=2} = \left(\bar{s}\gamma^\mu\frac{1-\gamma_5}{2}d\right)
              \left(\bar{s}\gamma^\mu\frac{1-\gamma_5}{2}d\right)
\equiv \bar{s}\gamma^\mu_L d \bar{s}\gamma^\mu_L d \;.
\end{equation}

We closely follow the RI procedure for renormalization of this operators
described in \cite{Donini:1995xj}. 
In the domain wall fermion formulation the
mixing with operators of different chiralities is small and is not
studied in our paper. For more detailed treatment of this topic
the reader is referred to Ref.~\cite{Dawson:2000kh}.
The four--point Green's function is defined by

\begin{equation}
  G_{B_K}(x_1,x_2,x_3,x_4) = \langle s(x_1)\bar{d}(x_2) O_{LL}^{\Delta{S}=2}(0)
                                   s(x_3)\bar{d}(x_4) \rangle \;.
\end{equation}

For the Fourier
transformed Green's function the off--shell external
momenta are chosen to be equal.
Using the upper Roman indices to denote color and the lower Greek indices
to denote spin, the non--amputated Green's function can be written as

\begin{equation}
  G_{B_K}(p)^{ABCD}_{\alpha\beta\gamma\delta} =
   2\left[ 
    \langle\Gamma^\mu(p)^{AB}_{\alpha\beta}  
        \Gamma^\mu(p)^{CD}_{\gamma\delta} \rangle
    -  \langle\Gamma^\mu(p)^{AD}_{\alpha\delta}  
        \Gamma^\mu(p)^{CB}_{\gamma\beta} \rangle
   \right] \;,
\end{equation}
where
\begin{equation}
  \Gamma^\mu(p)^{AB}_{\alpha\beta} = S(p|0)^{AR}_{\alpha\sigma}
         {\gamma^\mu_L}_{\sigma\rho}
         \left(\gamma_5 S(p|0)^\dag \gamma_5\right)^{RB}_{\rho\beta} \;.
\end{equation}
$S(p|0)$ is a Fourier transformed propagator on a single configuration
in Landau gauge. It is not translationally invariant.

The amputated Green's function is calculated from the non--amputated one
by multiplying it with the inverse propagator $S(p)$
averaged over all configurations
\begin{equation}
   \Lambda_{B_K}(p)^{ABCD}_{\alpha\beta\gamma\delta}
  = S^{-1}(p)^{AA'}_{\alpha\alpha'} S^{-1}(p)^{CC'}_{\gamma\gamma'}
      G_{B_K}(p)^{A'B'C'D'}_{\alpha'\beta'\gamma'\delta'}
   S^{-1}(p)^{B'B}_{\beta'\beta} S^{-1}(p)^{D'D}_{\delta'\delta} \;.
\end{equation}
The projection operation Eq.~(\ref{project}) is defined by the equation
\begin{equation}
  \Gamma_{B_K}(p;a) = 
    {1\over 32N_c(N_c+1)}
    {\gamma^\mu_L}_{\alpha\alpha'} {\gamma^\mu_L}_{\beta\beta'}
    \Lambda_{B_K}(p)^{AABB}_{\alpha'\alpha\beta'\beta} \;,
\end{equation}
where $N_c=3$ is the number of colors.
The renormalization condition imposed on the $O^{\Delta{S}=2}$ operator
is given by the same formula Eq.~(\ref{rencond}).

\begin{figure}
\epsfysize=20cm
\centering
\epsfbox{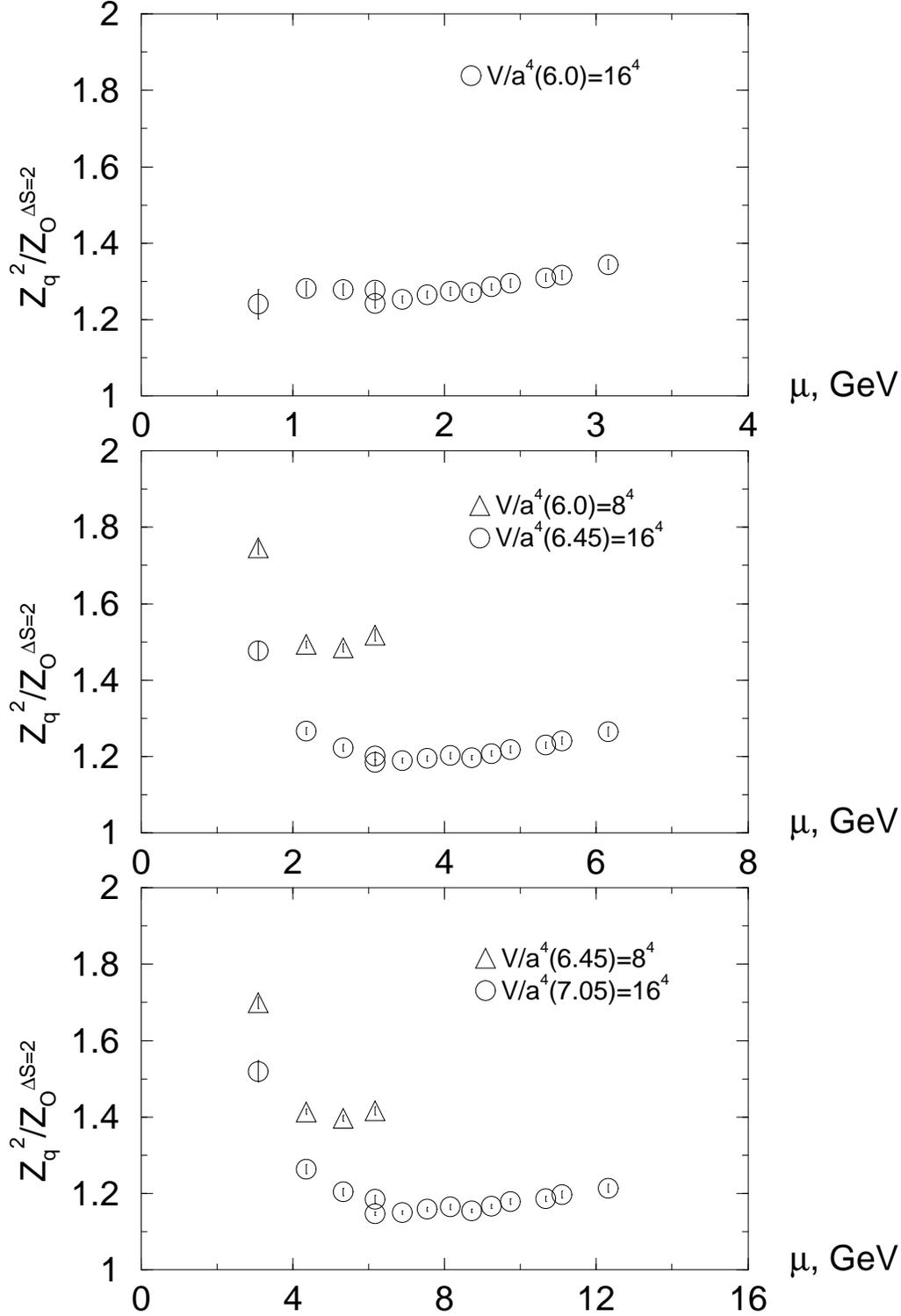}
\caption{$Z_q^2/Z_{O_{LL}^{\Delta S=2}}$ in the chiral limit
from local vector and axial currents. The graphs from top to
bottom correspond to different physical volumes, while each pair
of graphs is for the same physical volume.}
\label{fig:Zbk}
\end{figure}

\begin{figure}
\epsfxsize=15cm
\epsfbox{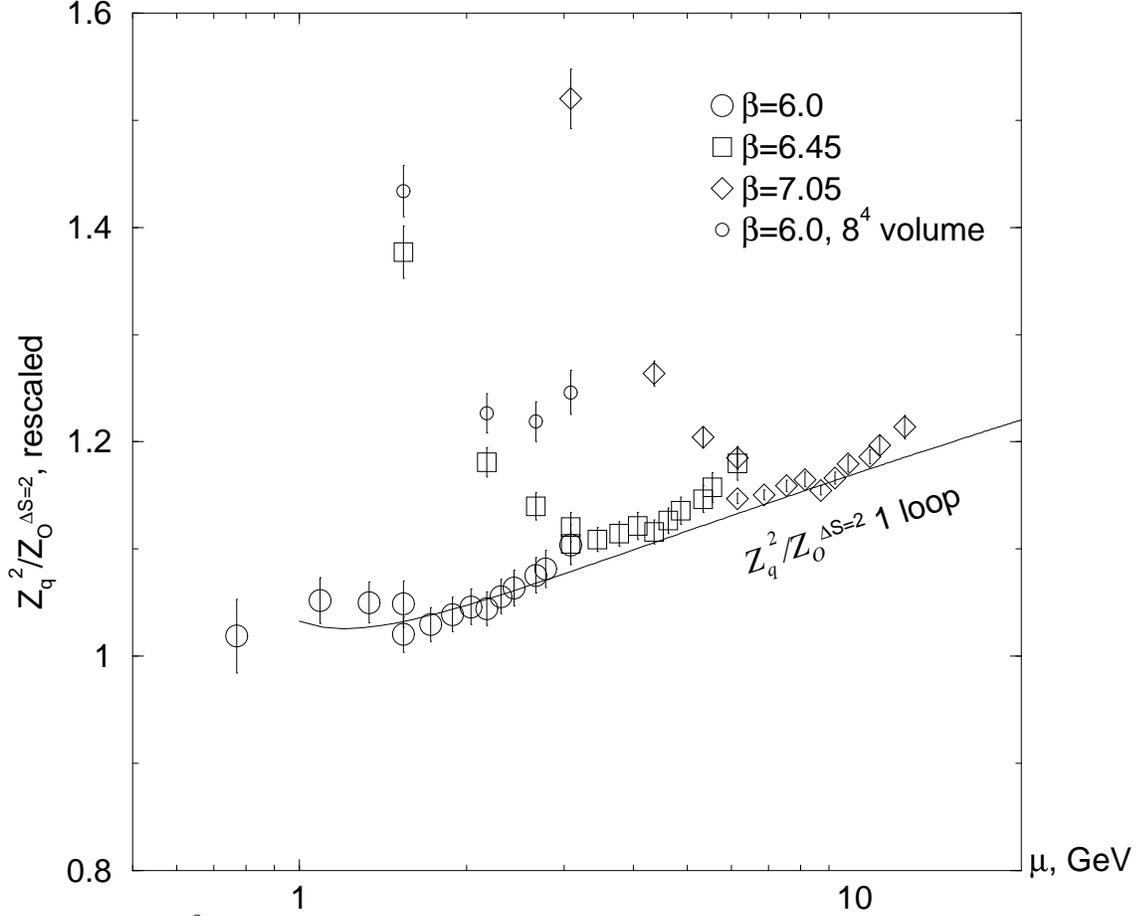}
\caption{$Z_q^2/Z_{O_{LL}^{\Delta S=2}}$ plotted for
 a wide range of scales $\mu$.
All graphs have the same ultraviolet cutoff equal to
the lattice spacing at $\beta=7.05$. 
 Different symbols correspond
to data in different physical volumes (labeled by the $\beta$ at which they are
obtained).
The finite volume effects are
large for momenta on the order of the inverse box size.
 Large symbols correspond to the $16^4$ lattice volume.
Small circles denote data obtained in a $8^4$ volume at $\beta=6.0$.
Since no relative renormalization is involved for the two
calculations at $\beta=6.0$, the difference between the points
is due to the finite volume effects only. The difference between
the small circles and squares is due to discretization errors
at the corresponding momenta. These errors are larger for 
the results obtained at the smaller $\beta=6.0$.}
\label{fig:ZbkAllbeta}
\end{figure}

The numerical results for this $Z$--factor obtained with different
values of $m_f$ have no statistically significant dependence 
on $m_f$. Nevertheless, the same procedure of linear correlated
fitting in $m_f$ was applied to take the chiral limit with
the results shown in Figure \ref{fig:Zbk}.
The final results after rescaling to a new ultraviolet cutoff
equal to the lattice spacing at $\beta=7.05$ are
 presented in Figure~\ref{fig:ZbkAllbeta}.
It is interesting to notice that the finite volume effects at
low lattice momenta are very significant and for the lowest momenta
are of the order of $50$\%. This same volume dependence can
also be seen from a direct comparison with the smaller, $8^4$ volume
at the same $\beta$. For illustration purposes we include the
result at $\beta=6.0$ in a $8^4$ volume in Figure~\ref{fig:ZbkAllbeta}
using small circles.
The data points for this smaller volume are rescaled using the same factor
as the $\beta=6.0$, $16^4$ volume result, therefore the difference
between the two results is a direct result of the finite volume effects.
The systematic difference
from the data at $\beta=6.45$ that has the same physical volume is
due to larger discretization errors that are present in the $\beta=6.0$
data at the corresponding momenta.

The solid line on the graph is for the one--loop renormalization
group analysis in perturbation theory \cite{Ciuchini:1998bw}.
 Since it is defined up to
an overall multiplicative constant which we do not determine,
we matched it with the 
numerical result at large energy scales $\mu$ of about $10$ GeV.
There is a surprisingly good agreement in the whole range from $1$ to $10$ GeV.
This might be fortuitous
 for $\mu$ lower than $2$ GeV, since at these scales
we expect large non--perturbative effects. 
These effects together with the large 
finite volume effects at low momenta could have resulted in the flat shape
of numerical results at around $1$ GeV.

\section{Conclusions}

The combination of recursive
scaling and NPR offers a promising connection between the perturbative
and non--perturbative regimes--a connection that is required if
perturbative errors are to be properly controlled. 
The results reported here suggest a consistent picture for the
scaling of renormalization factors over a momentum range between 2
and 10 GeV for the case of quenched domain wall fermions for the
operators analyzed.  The only exception to this generally satisfactory
picture occurs in the comparison of the local and conserved currents,
where the larger discretization errors present in the non--local,
conserved currents may be at fault.
To fully resolve this issue and demonstrate the validity
of this approach more extensive
calculations and larger lattice volumes will be required.

\section{Acknowledgments}

The author thanks C.~Dawson for much important assistance with
the NPR technique and N.~Christ, T.~Blum, and R.~Mawhinney
for helpful discussions, and other collaborators and Columbia
and Brookhaven for the hardware and software resources which
were essential to this work.

All of the numerical calculations were done on the 400 Gflops QCDSP 
computer at Columbia University and the 600 Gflops QCDSP computer at
the RIKEN--BNL Research Center.
This work was supported in part by the U.S. Department of Energy and
the RIKEN BNL Research Center.


\bibliography{lattice,thermo}



\begin{table*}[htb]
\caption{Rescaling coefficients.}
\label{table:1}
\begin{tabular}{||l|c|c||}
$Z$--factor & $R(a(\beta=6.45),a(\beta=6.0))$ 
	& $R(a(\beta=7.05),a(\beta=6.45))$ \\ \hline
$Z_q/Z_{V^L}$ & 1.016(3) & 0.999(2) \\
$Z_q/Z_{A^L}$ & 1.014(3) & 0.987(2) \\
$Z_q/Z_P$ & 1.088(9) & 1.039(6) \\
$Z_q/Z_S$ & 1.046(8) & 1.025(6) \\
$Z_q^2/Z_{O^{\Delta S=2}_{LL}}$ & 0.881(9) & 0.932(8) \\
$Z_q$  & 1.188(13) & 1.067(5) \\
$Z^{(A)}_q$ & 1.092(13)& 1.047(5) \\
\end{tabular}
\end{table*}

\begin{table*}[htb]
\caption{$Z$--factors at $\beta=6.0$ in $16^4$ lattice volume.
The mass for the local $Z$--factors is $m_f=0.02$, while for 
the conserved currents
(last two columns) $m_f=0.05$.}
\label{table:2}
\begin{tabular}{||c|c||c|c|c|c|c||c|c||}
$p_i a$&$\mu$,GeV&${Z_q/Z_{V^L}}$&${Z_q/Z_{A^L}}$
&${Z_q/Z_P}$&${Z_q/Z_S}$&$Z_q^2/Z_{O_{LL}^{\Delta S=2}}$
&$Z_q$&$Z^{(A)}_q$\\ \hline 
0001 & 0.770 & 1.1859(91) & 0.9865(71) & 5.610(22) 
& 1.43(14) & 1.229(25) & 0.914(16) & 0.748(12) \\
0011 & 1.089 & 1.1350(48) & 1.0545(39) & 3.620(13) 
& 1.430(89)  & 1.258(15) & 0.862(13)  & 0.790(11) \\
0111 & 1.333 & 1.1142(41) & 1.0731(35) & 2.799(89) 
& 1.451(65) & 1.265(13)& 0.841(12)  & 0.801(12) \\
0002 & 1.539 & 1.0882(27) & 1.0660(26) & 2.361(66) 
& 1.466(50) & 1.238(10) & 0.824(14) & 0.800(14) \\
1111 & 1.539 & 1.1034(45) & 1.0820(45) & 2.314(54) 
& 1.456(37) & 1.273(15) & 0.828(11) & 0.802(10) \\
0012 & 1.721 & 1.0915(23) & 1.0749(21) & 2.081(48) 
& 1.432(36) & 1.249(9) & 0.836(10) & 0.817(10) \\
0112 & 1.885 & 1.0917(25) & 1.0804(23) & 1.907(36) 
& 1.419(25) & 1.261(9)& 0.841(11)  & 0.828(11) \\
1112 & 2.036 & 1.0927(29) & 1.0836(29) & 1.785(28) 
& 1.405(18) & 1.271(10)  & 0.844(12) & 0.836(12)\\
0022 & 2.177 & 1.0913(22) & 1.0835(21) & 1.703(24) 
& 1.386(16) & 1.268(8)& 0.873(10)  & 0.863(10) \\
0122 & 2.309 & 1.0938(24) & 1.0880(24) & 1.639(22) 
& 1.370(15) & 1.282(9) & 0.888(10) & 0.881(10) \\
1122 & 2.434 & 1.0962(29) & 1.0913(29) & 1.591(18)
& 1.364(12) & 1.292(10)  & 0.902(11) & 0.897(11) \\
0222 & 2.666 & 1.1004(27) & 1.0966(27) & 1.522(16) 
& 1.339(12) & 1.307(10) & 0.958(11) & 0.953(11) \\
1222 & 2.775 & 1.1025(32) & 1.0993(32) & 1.497(13) 
& 1.344(10) & 1.315(11) & 0.977(12)& 0.971(12)  \\
2222 & 3.079 & 1.1100(36) & 1.1079(36) & 1.441(10) 
& 1.328(09) & 1.342(13)  & 1.057(14) & 1.053(14) \\
\end{tabular}
\end{table*}

\begin{table*}[htb]
\caption{Rescaled $Z$--factors in the chiral limit versus the energy scale. 
The $Z$--factors in this table are defined with the ultraviolet cutoff
equal to the lattice spacing at $\beta=7.05$. The physical extent
of the lattice in each dimension is 16 times the lattice
spacing at $\beta=6.0$.}
\label{table:3}
\begin{tabular}{||c||c|c|c|c|c||c|c||}
$\mu$,GeV&${Z_q/Z_{V^L}}$&${Z_q/Z_{A^L}}$
&${Z_q/Z_P}$&${Z_q/Z_S}$&$Z_q^2/Z_{O_{LL}^{\Delta S=2}}$
&$Z_q$&$Z^{(A)}_q$\\ \hline 
0.770 & 1.1910(15) & 1.0045(12) & 2.1989(147) & 2.0848(78) 
& 1.0188(34) & 1.1591(24) & 0.8554(18) \\
1.089 & 1.1533(8) & 1.0691(7) & 1.9869(79) & 1.8985(47) 
& 1.0521(21) & 1.0939(21) & 0.9034(17) \\
1.333 & 1.1309(7) & 1.0831(6) & 1.8653(55) & 1.7918(37) 
& 1.0500(19) & 1.0668(20) & 0.9167(18) \\
1.539 & 1.1014(5) & 1.0726(5) & 1.7911(38) & 1.7010(29) 
& 1.0202(17) & 1.0447(22) & 0.9153(20) \\
1.539 & 1.1200(7) & 1.0903(7) & 1.8335(40) & 1.7317(28) 
& 1.0487(22) & 1.0500(19) & 0.9171(17) \\
1.721 & 1.1069(5) & 1.0815(5) & 1.7470(31) & 1.6485(23) 
& 1.0294(16) & 1.0598(18) & 0.9347(17) \\
1.885 & 1.1076(5) & 1.0857(5) & 1.6891(27) & 1.6035(21) 
& 1.0389(16) & 1.0662(18) & 0.9466(17) \\
2.036 & 1.1082(5) & 1.0864(5) & 1.6466(24) & 1.5668(21) 
& 1.0459(17) & 1.0707(20) & 0.9563(18) \\
2.177 & 1.1074(5) & 1.0880(5) & 1.6230(21) & 1.5363(18) 
& 1.0445(16) & 1.1067(18) & 0.9870(17) \\
2.309 & 1.1098(5) & 1.0916(5) & 1.5988(20) & 1.5142(17) 
& 1.0557(16) & 1.1268(19) & 1.0076(17) \\
2.434 & 1.1127(5) & 1.0942(5) & 1.5776(20) & 1.4973(17) 
& 1.0636(17) & 1.1446(20) & 1.0255(18) \\
2.666 & 1.1169(5) & 1.0989(5) & 1.5500(18) & 1.4673(16) 
& 1.0755(17) & 1.2148(20) & 1.0895(18) \\
2.775 & 1.1186(5) & 1.1013(5) & 1.5393(18) & 1.4635(16) 
& 1.0812(17) & 1.2391(21) & 1.1108(20) \\
3.079 & 1.1268(6) & 1.1103(6) & 1.5212(18) & 1.4394(16) 
& 1.1040(18) & 1.3404(23) & 1.2039(22) \\
\end{tabular}
\end{table*}

\begin{table*}[htb]
\caption{Rescaled $Z$--factors in the chiral limit versus the energy scale. 
The $Z$--factors in this table are defined with the ultraviolet cutoff
equal to the lattice spacing at $\beta=7.05$. The physical extent
of the lattice in each dimension is 16 times the lattice
spacing at $\beta=6.45$.}
\label{table:4}
\begin{tabular}{||c||c|c|c|c|c||c|c||}
$\mu$,GeV&${Z_q/Z_{V^L}}$&${Z_q/Z_{A^L}}$
&${Z_q/Z_P}$&${Z_q/Z_S}$&$Z_q^2/Z_{O_{LL}^{\Delta S=2}}$
&$Z_q$&$Z^{(A)}_q$\\ \hline 
1.539 & 1.1052(6) & 1.0861(6) & 2.2889(33) & 2.0970(37) & 1.3771(25) 
& 0.9897(17) & 0.8713(13) \\
2.177 & 1.0782(3) & 1.0634(3) & 1.7676(18) & 1.7055(19) & 1.1810(14) 
& 0.9178(12) & 0.8718(11) \\
2.666 & 1.0699(3) & 1.0553(3) & 1.5857(14) & 1.5470(15) & 1.1400(13) 
& 0.8827(12) & 0.8522(11) \\
3.079 & 1.0595(3) & 1.0457(3) & 1.4952(13) & 1.4609(13) & 1.1045(11) 
& 0.8665(12) & 0.8419(12) \\
3.079 & 1.0654(4) & 1.0502(4) & 1.5136(13) & 1.4818(13) & 1.1204(14) 
& 0.8709(10) & 0.8462(10) \\
3.442 & 1.0596(3) & 1.0458(3) & 1.4524(11) & 1.4244(11) & 1.1090(11) 
& 0.8665(9) & 0.8442(9) \\
3.771 & 1.0609(3) & 1.0478(3) & 1.4101(10) & 1.3840(10) & 1.1142(12) 
& 0.8667(10) & 0.8461(10) \\
4.073 & 1.0635(4) & 1.0504(3) & 1.3793(10) & 1.3586(10) & 1.1215(13) 
& 0.8719(11) & 0.8520(11) \\
4.354 & 1.0623(3) & 1.0485(3) & 1.3663(9) & 1.3415(10) & 1.1160(11) 
& 0.8797(9) & 0.8603(9) \\
4.618 & 1.0654(3) & 1.0517(3) & 1.3484(9) & 1.3261(9) & 1.1264(12) 
& 0.8906(10) & 0.8717(9) \\
4.868 & 1.0684(4) & 1.0550(4) & 1.3336(9) & 1.3135(9) & 1.1358(13) 
& 0.9050(11) & 0.8862(11) \\
5.333 & 1.0717(3) & 1.0580(3) & 1.3150(9) & 1.2969(9) & 1.1466(13) 
& 0.9365(10) & 0.9176(10) \\
5.550 & 1.0751(4) & 1.0613(4) & 1.3110(9) & 1.2904(9) & 1.1575(14) 
& 0.9572(12) & 0.9382(12) \\
6.158 & 1.0823(4) & 1.0688(4) & 1.2990(8) & 1.2806(9) & 1.1798(15) 
& 1.0215(13) & 1.0018(13) \\
\end{tabular}
\end{table*}

\begin{table*}[htb]
\caption{Rescaled $Z$--factors in the chiral limit versus the energy scale. 
The $Z$--factors in this table are defined with the ultraviolet cutoff
equal to the lattice spacing at $\beta=7.05$. The physical extent
of the lattice in each dimension is 16 times the lattice
spacing at $\beta=7.05$.}
\label{table:5}
\begin{tabular}{||c||c|c|c|c|c||c|c||}
$\mu$,GeV&${Z_q/Z_{V^L}}$&${Z_q/Z_{A^L}}$
&${Z_q/Z_P}$&${Z_q/Z_S}$&$Z_q^2/Z_{O_{LL}^{\Delta S=2}}$
&$Z_q$&$Z^{(A)}_q$\\ \hline 
3.079 & 1.0748(8) & 1.0624(4) & 1.8674(28) & 1.8050(30) & 1.5200(28) 
& 0.8407(12) & 0.8100(12) \\
4.354 & 1.0630(2) & 1.0604(2) & 1.4970(13) & 1.4777(13) & 1.2637(12) 
& 0.8324(8) & 0.8247(8) \\
5.333 & 1.0593(2) & 1.0586(2) & 1.3698(9) & 1.3612(9) & 1.2041(10) 
& 0.8074(8) & 0.8032(8) \\
6.158 & 1.0416(1) & 1.0406(1) & 1.3058(8) & 1.3017(8) & 1.1472(5) 
& 0.7888(7) & 0.7848(7) \\
6.158 & 1.0571(3) & 1.0572(3) & 1.3208(7) & 1.3176(7) & 1.1850(11) 
& 0.7958(8) & 0.7931(8) \\
6.884 & 1.0452(1) & 1.0442(1) & 1.2806(5) & 1.2752(5) & 1.1504(5) 
& 0.7883(7) & 0.7860(7) \\
7.541 & 1.0494(2) & 1.0487(2) & 1.2527(5) & 1.2492(5) & 1.1589(5) 
& 0.7854(6) & 0.7837(6) \\
8.146 & 1.0518(2) & 1.0517(2) & 1.2320(5) & 1.2291(5) & 1.1649(7) 
& 0.7877(7) & 0.7862(7) \\
8.708 & 1.0483(1) & 1.0475(1) & 1.2238(4) & 1.2199(4) & 1.1546(4) 
& 0.7885(6) & 0.7872(6) \\
9.236 & 1.0525(2) & 1.0519(2) & 1.2127(4) & 1.2100(4) & 1.1663(6) 
& 0.7954(6) & 0.7942(6) \\
9.736 & 1.0562(2) & 1.0563(2) & 1.2033(4) & 1.2012(4) & 1.1794(8) 
& 0.8053(7) & 0.8043(7) \\
10.665 & 1.0590(2) & 1.0587(2) & 1.1941(4) & 1.1925(4) & 1.1861(7) 
& 0.8247(6) & 0.8239(6) \\
11.101 & 1.0622(3) & 1.0621(3) & 1.1896(4) & 1.1882(4) & 1.1969(9) 
& 0.8400(7) & 0.8392(7) \\
12.315 & 1.0678(3) & 1.0676(3) & 1.1806(4) & 1.1790(4) & 1.2139(11) 
& 0.8831(8) & 0.8824(8) \\
\end{tabular}
\end{table*}

%

\end{document}